\documentclass[preprint, 12pt]{elsarticle}

\usepackage{amssymb}

\usepackage{color}
\usepackage{graphicx}
\usepackage{subfigure}
\usepackage{amsmath}
\usepackage{bm}

\newcounter{bla}

\journal{Elsevier}

\begin{document}

\begin{frontmatter}



\title{Dynamics of a droplet in shear flow by smoothed particle hydrodynamics}


\author[a]{Kuiliang Wang}
\author[b]{Hong Liang}
\author[c]{Chong Zhao}
\author[a]{Xin Bian\corref{author}}

\cortext[author] {
Corresponding author.\\\textit{E-mail address:} bianx@zju.edu.cn
}
\address[a]{
State Key Laboratory of Fluid Power and Mechatronic Systems, Department of Engineering Mechanics, Zhejiang University, Hanghzhou 310027, China.
}
\address[b]{
Department of Physics, Hangzhou Dianzi University, Hangzhou 310018, China.
}
\address[c]{
Hangzhou Shiguangji Intelligient Electronics Technology Co., Ltd, Hangzhou, 310018, China.
}
\begin{abstract}
We employ a multi-phase smoothed particle hydrodynamics (SPH) method to study droplet dynamics in shear flow. With an extensive range of Reynolds number, capillary number, wall confinement, and density/viscosity ratio between the droplet and the matrix fluid, we are able to investigate systematically the droplet dynamics such as deformation and breakup. 
We conduct the majority of the simulations in two dimensions due to economical computations, while perform a few representative simulations in three dimensions to corroborate the former. 
Comparison between current results and those in literature indicates that the SPH method adopted has an excellent accuracy and is capable of simulating scenarios with large density or/and viscosity ratios. We generate slices of phase diagram in five dimensions, scopes of which are unprecedented. Based on the phase diagram, critical capillary numbers can be identified on the boundary of different states. As a realistic application, we perform simulations with actual parameters of water droplet in air flow to predict the critical conditions of breakup, which is crucial in the context of atomization.

\end{abstract}


\end{frontmatter}

\section{Introduction}
\label{sec:introduction}

 The deformation and breakup of droplets in shear flow are ubiquitous in engineering applications. 
On microfluidic chips, droplets are utilized for microbial cultivation and material transport~\cite{anna2016droplets,dressler2017chemical},
and a thorough understanding of their dynamics in confined flows may improve the efficiency of production and transportation.  
In other environmental and industrial applications such as protection against harmful aerosols, ink-jet printing and atomization in nozzles~\cite{liu2022numerical, lohse2022fundamental, aydin2011experimental, si2014design, xu2020droplet},
liquid droplets are typically in gas flows.
Accordingly, a decent knowledge on their dynamics 
with a high density/viscosity ratio against the matrix fluid is significant.
To this end, a comprehensive investigation on the dynamics of a droplet in shear flow, which involves a wide range of Reynolds number, capillary number, confinements of the wall, viscosity/density ratio between the two phases, is called for.

Since pioneering works by Taylor on droplet deformation in shear and extensional flows~\cite{taylor1932ViscosityFluidContaining, taylor1934FormationEmulsionsDefinable}, enormous theoretical and experimental studies have been conducted. 
A series of works by the group of Mason~\cite{bartok1959particle, rumscheidt1961particle,torza1971particle} further studied the deformation and burst of droplets, and even depicted the streamlines inside and around the droplets. 
Chaffey and Brenner~\cite{chaffey1967second} extended a previous analytical approximation to a second order form, which is crucial for the non-elliptic deformation of a highly viscous droplet under large shear rate. 
Barthes-Biesel and Acrivos~\cite{barthes1973deformation} expressed the solution of creeping-flow equations in powers of deformation parameters and applied a linear stability theory to determine the critical values for the droplet breakup.
Hinch and Acrivos~\cite{hinch1980long} investigated theoretically the stability of a long slender droplet, which is largely deformed in shear flow. 
However, early analytical works rarely considered effects of finite Reynolds number or wall confinements.
In addition, numerous experimental studies have been conducted on the droplet deformation and  breakup~\cite{karam1968deformation, flumerfelt1972drop, stone1986experimental, guido1998three},
where not only the effects of viscosity ratio between the droplet and the matrix fluid~\cite{grace1982dispersion, stone1989influence}, but also wall confinements~\cite{vananroye2006effect, vananroye2007effect} have been taken into account.
 
With advance in computational science, numerical simulation has become a popular approach to study droplet dynamics in the past decades. Boundary integral method was among the first to be applied to study deformation of droplets in stationary and transient states~\cite{kennedy1994motion}, non-Newtonian droplets~\cite{toose1995boundary}, and migration of a droplet in shear flow~\cite{uijttewaal1995motion}. Moreover, Li et al.~\cite{li2000numerical} employed a volume-of-fluid (VOF) method and Galerkin projection technique to simulate the process of droplet breakup. In the work of Amani et al.~\cite{amani2019numerical}, a conservative level-set (CLS) method built on a conservative finite-volume approximation is applied to study the effect of viscosity ratio and wall confinement on the critical capillary number. In addition, lattice Boltzmann method (LBM) has been widely employed to study deformation, breakup and coalescence of droplets~\cite{xi1999lattice, van2008emulsion, farokhirad2013effects, komrakova2014lattice, huang2022lattice}; to model viscoelastic droplet~\cite{wang2020lattice} and surfactant-laden droplet~\cite{zong2020modeling}.
We note that an interface tracing technique such as VOF, CLS, a phase-field formulation,
or immersed boundary method is often necessary by a flow solver based on Eulerian meshes.

As a Lagrangian method, smoothed particle hydrodynamics (SPH) method has some advantages in simulating multiphase flows. Since different phases are identified by different types of particles, the interface automatically emerges without an auxillary tracing technique, even for a very large deformation.
Moreover, inertia and wall effects can  be taken into account straightforward, in contrast to theoretical analysis or the boundary integral method.  
Since its inception in astrophysics, SPH method has been largely developed and widely applied in various flow problems~\cite{Monaghan2012, ye2019smoothed}. Morris~\cite{morris2000SimulatingSurfaceTension} considered the surface tension based on a continuous surface force model and simulated an oscillating two-dimensional rod in SPH. Hu et al.~\cite{hu2006MultiphaseSPHMethod} proposed a multi-phase model that handles both macroscopic and mesoscopic flows in SPH, where a droplet in shear flow was selected as a benchmark to validate the method. Other improvements and modifications have also been proposed for SPH in the context of multiphase problems~\cite{ wang2016overview, zhang2010simulation, tartakovsky2016pairwise, yang2019comprehensive}.
Furthermore, a droplet or matrix flow with special properties can also be considered. For example, Moinfar et al.~\cite{moinfar2022numerical} studied the drop deformation under simple shear flow of Giesekus fluids and Vahabi~\cite{vahabi2022effect} investigated the effect of thixotropy on deformation of a droplet under shear flow. Saghatchi et al.~\cite{saghatchi2021dynamics} studied the dynamics of a 2D double emulsion in shear flow with electric field based on an incompressible SPH method.
There are also studies on colliding and coalescence process of droplets by SPH~\cite{hirschler2017modeling, xu2020modified}.
Simulation of bubbles in liquid is similar, but can encounter special challenges~\cite{zhang2015sph}, due to the reverse density/viscosity ratio as that of droplet in gas. 

Previously, simulations of multiphase flows by SPH method often investigated specific circumstances. Therefore, the objective of this paper is two fold: firstly, to simulate an extensive range of parameters to examine the SPH method for multiphase flows;
secondly, to fill gaps of unexplored range of parameters and systematically investigate their influence on the droplet dynamics.
The rest of the paper is arranged as follows: in Sec.~\ref{sec:method}, we introduce the multiphase SPH method and a specific surface tension model. We present validations and extensive numerical results in Sec.~\ref{sec:results}. We summarize this work after discussions in Sec.~\ref{sec:conclusion}.

\section{Method}
\label{sec:method}

\subsection{Governing equations and surface tension model}
We consider isothermal Navier-Stokes equations with a surface tension for multiphase flow in Lagrangian frame
\begin{equation}
\begin{aligned}
&\frac{d\rho }{dt} =-\rho\nabla \cdot \mathbf{v}, \\
&\frac{d\mathbf{v}}{dt} = \frac{1}{\rho}  \left ( -\nabla p + \mathbf{F}_{b} + \mathbf{F}_{v} + \mathbf{F}_{s} \right ),
\end{aligned}
 \label{governing}
\end{equation}
where $\rho$, $\mathbf{v}$ and $p$ are density, velocity and pressure respectively. $\mathbf{F}_{b}$ is the body force, which is not considered in this study.  $\mathbf{F}_{v}$, $\mathbf{F}_{s}$ denote viscous force and surface tension at the interface between two phases, respectively.

Following previous studies of quasi-incompressible flow modeling~\cite{morris2000SimulatingSurfaceTension}, an artificial equation of state relating pressure to density can be written as 
\begin{equation}
p=c_{s}^{2} \left ( \rho - \rho_{\mathrm{ref}} \right ),
\label{state}
\end{equation}
where $c_{s}$ is an artificial sound speed and $\rho_{\mathrm{ref}}$ is a reference density. Theoretically, subtracting the reference density has no influence on the gradient of pressure, but it can reduce the numerical error 
of SPH discretizations for the gradient operator.

For a Newtonian flow, the viscous force $\mathbf{F}_{v}$ simplifies to
\begin{equation}
\mathbf{F}_{v}=\mu\nabla^{2}\mathbf{v},
\label{visco_force}
\end{equation}
where $\mu$ is the dynamic viscosity. 
We assume surface tension to be uniform along the interface
and do not consider Marangoni force.
Therefore, the surface tension acts on the normal direction of the interface.
Moreover, its magnitude depends on the local curvature as
\begin{equation}
\mathbf{F} _{s}=\sigma \kappa \mathbf{\hat{n}}\delta_{s},
\label{surface_tension1}
\end{equation}
where $\sigma$, $\kappa$, $\mathbf{\hat{n}}$ are surface tension coefficient, curvature and unit normal vector to the concave side, respectively; $\delta_{s}$ is a surface delta function and its discrete form shall be described later.

To describe the surface tension at the interface between two fluids, a continuous surface tension model is adopted. 
As a matter of fact, surface tension my be written as the divergence of a tensor $\mathbf{T}$~\cite{brackbill1992continuum, lafaurie1994modelling}
\begin{equation}
\sigma \kappa \mathbf{\hat{n}}\delta_{s}=  \nabla \cdot \mathbf{T},
\label{surface_tensor1}
\end{equation}
where
\begin{equation}
\mathbf{T} = \sigma  \left ( \mathbf{I} - \mathbf{\hat{n}}  \otimes  \mathbf{\hat{n}}\right )\delta_{s}.
\label{surface_tensor2}
\end{equation}

To represent a multiphase flow,
we define a color function $c$ and set a unique value for each phase,
that is, $c^\mathrm{I}=0$ and $c^\mathrm{II}=1$ for the two phases, respectively.
Apparently, the color function has a jump from $0$ to $1$ at the interface between phase $\mathrm{I}$ and $\mathrm{II}$. 
Therefore, the unit normal vector can be represented by the normalized gradient of the color function as
\begin{equation}
\mathbf{\hat{n}}=\frac {\nabla c}{\left | \nabla c \right |},
\label{normal_gradient1}
\end{equation}
and the surface delta function is replaced by the scaled gradient as
\begin{equation}
\delta_{s} = \left | \mathbf{n} \right |=\frac {\left |\nabla c \right |}{\left | c^\mathrm{I}-c^\mathrm{II} \right |}.
\label{normalized_n}
\end{equation}
\subsection{SPH method}
In SPH, fluid is represented by moving particles carrying flow properties
such as density, velocity and pressure. 
We largely follow the work of Hu and Adams~\cite{hu2006MultiphaseSPHMethod}
and provide a brief derivation here.
Density of a particle is calculated by interpolating the mass of neighboring particles as
\begin{equation}
\rho_{i}=m_{i} \sum_{j}W_{ij},
\label{rhosum}
\end{equation}
where mass $m_i$ is constant for every particle. 
$W_{ij}$ denotes a weight function for interpolation
\begin{equation}
W_{ij}=W\left ( \mathbf{r} _{ij},h \right ),
\label{kernel}
\end{equation}
where $\mathbf{r} _{ij}=\mathbf{r} _{i}-\mathbf{r} _{j}$ is a relative position vector from particle $j$ to $i$ and $h$ is the smoothing length. We further define 
\begin{equation}
V_i = \frac{1}{\sum_{j}W_{ij}},
\label{equivalent_volume}
\end{equation}
to be an equivalent volume of particle $i$ so that $V_i = m_i / \rho_i $.

The pressure gradient can be computed as 
\begin{equation}
-\left ( \frac{1}{\rho} \nabla p \right ) _i = - \sum_{j}\left ( V_{i}^{2}p_i + V_{j}^{2}p_j \right ) \frac{\partial W}{\partial r_{ij}} \mathbf{e}_{ij},
\label{pressure_sph}
\end{equation}
where $p_i$ and $p_j$ are obtained by Eq.~(\ref{state}). The viscous force can be calculated as 
\begin{equation}
\left(\mu\nabla^{2}\mathbf{v}\right ) _i=\sum_{j}\frac{2\mu_i \mu_j}{\mu_i + \mu_j} \left ( V_i^2 + V_j^2 \right ) \frac{\mathbf{v}_{ij}}{r_{ij}} \frac{\partial W}{\partial r_{ij}},
\label{visco_sph}
\end{equation}
where $\mathbf{v}_{ij}=\mathbf{v}_{i} - \mathbf{v}_{j}$ is the relative velocity of particle $i$ and $j$ and $r_{ij}=\left | \mathbf{r} _{ij} \right |$ is the distance between them.

As suggested by Morris~\cite{morris2000SimulatingSurfaceTension} and Hu et al.~\cite{hu2006MultiphaseSPHMethod}, a part of pressure contribution $\sigma \frac{d-1}{d} \delta_{s}$ is removed to avoid attractive force and improve the stability of the interactions between SPH particles. Therefore, we employ
\begin{equation}
\mathbf{T}' = \sigma  \left (\frac{1}{d} \mathbf{I} - \mathbf{\hat{n}}  \otimes  \mathbf{\hat{n}}\right )\delta_{s}
\label{surface_tensor3}
\end{equation}
to replace Eq.~(\ref{surface_tensor2}), where $d$ is the spatial dimension. Combining Eq.~(\ref{normalized_n}), Eq.~(\ref{normal_gradient1}) and Eq.~(\ref{surface_tensor3}), we obtain
\begin{equation}
\mathbf{T}'=\frac{\sigma }{\left | c^\mathrm{I}-c^\mathrm{II} \right | \left | \nabla c \right | } \left ( \frac{\left | \nabla c \right |^2 }{d} \mathbf{I} - \nabla c \otimes \nabla c \right ).
\label{surface_tensor4}
\end{equation}
The gradient of color function between phase $\mathrm{I}$ and phase $\mathrm{II}$ can be calculated in SPH as
\begin{equation}
\nabla c_i = \frac{1}{V_i } \sum_{j}V_j^2\left ( c_j - c_i\right )\frac{\partial W}{\partial r_{ij}} \mathbf{e}_{ij},
\label{color_gradient_sph}
\end{equation}
where $c_i$ (or $c_j$) is initially assigned to be $c^\mathrm{I}$ or $c^\mathrm{II}$
according to which phase particle $i$ (or $j$) consititutes. 
Substitute Eq.~(\ref{color_gradient_sph}) into Eq.~(\ref{surface_tensor4}) to obtain stress tensor
\begin{equation}
\mathbf{T}'_i=\frac{\sigma }{ \left |  \nabla c_i \right | \left | c^{\mathrm {I}}-c^{\mathrm {II}} \right | } \left ( \frac{\left | \nabla c_i \right |^2 }{d} \mathbf{I} - \nabla c_i \otimes \nabla c_i \right ).
\label{stress_tensor_sph}
\end{equation}
Finally, the surface force term is calculated by the stress tensor using the SPH expression for divergence
\begin{equation}
\left( \sigma \kappa \mathbf{\hat{n}}\delta_{s} \right)_i = \sum_{j}\frac{\partial W}{\partial r_{ij}} \mathbf{e}_{ij} \cdot \left ( V_i^2\mathbf{T}'_i + V_j^2\mathbf{T}'_j  \right ).
\label{surface_tensor_sph}
\end{equation}
It is simple to see that the discrete version of $\delta_s$ in SPH is 
\begin{equation}
    \left( \delta_s \right)_i = \frac{1}{V_i \left | c^{\mathrm {I}}-c^{\mathrm {II}} \right | } \left | \sum_{j}V_j^2\left ( c_j - c_i\right )\frac{\partial W}{\partial r_{ij}} \mathbf{e}_{ij} \right |,
\end{equation}
which has a finite support to remove the singularity and distributes the surface tension onto a thin layer of two fluids across the interface.

\subsection{Computational settings}
The quintic kernel is adopted as weight function
\begin{equation}
W =\phi \begin{cases}
  (3-R)^5-6(2-R)^5+15(1-R)^5 & 0 \le R < 1; \\
  (3-R)^5-6(2-R)^5 & 1 \le R < 2; \\
  (3-R)^5 & 2 \le R < 3; \\
  0 & R \ge 3,
\end{cases}
\label{quintic}
\end{equation}
where $R=r/h$ and $h$ is the smoothing length. $\phi$ is a normalization coefficient which equals $1/120$, $7/(478\pi)$ and $1/(120\pi)$ in one, two and three dimensions, respectively. We set $h=1.2\Delta x$ with $\Delta x$ as the initial spacing distance between particles. This means that the support domain of the kernel function is truncated at $3.6\Delta x$, namely the cutoff $r_c = 3.6\Delta x$. According to our tests, a smoothing length of $1.2\Delta x$ is almost optimal for an excellent accuracy while avoiding the pairing instability. A detailed discussion on this issue is referred to Price~\cite{price2012SmoothedParticleHydrodynamics}.

Since we adopt a weakly compressible formulation, 
the sound speed $c_s$ should be large enough to restrict the density fluctuations.
Based on a scale analysis, Morris et al.~\cite{morris2000SimulatingSurfaceTension,morris1997ModelingLowReynolds} suggested that $c^2_s$ should be comparable to the largest of
\begin{equation}
 \frac{U^2}{\Delta } ,\ \frac{\mu U}{\rho_0 L \Delta} ,\  \frac{FL}{\Delta} ,\  \frac{\sigma \kappa }{\rho_0\Delta},
\label{cs_chosen}
\end{equation}
where $\Delta$ is the density variation and $U$, $L$, $F$, $\kappa$ and $\sigma$ are typical velocity, length, body force, curvature and surface tension coefficient, respectively. 
Accordingly, for multiphase flows the sound speed may be different
for each phase.
In all simulations, we set identical $\Delta \le 0.5\%$ for each phase and calculate $c_s$ accordingly. 

At every time step, the minimal relative density is recorded among all particles, that is,
\begin{equation}
\rho_{min}=min\left \{ min\left \{ \frac{\rho _i}{\rho _{0}^{\mathrm {I}} }  \right \}, min\left \{ \frac{\rho _j}{\rho _{0}^{\mathrm {II}} }  \right \}  \right \},
\label{rho_min}
\end{equation}
where particle $i$ belongs to phase $\mathrm {I}$ and particle $j$ belongs to phase $\mathrm {II}$; $\rho_0^{\mathrm {I}}$, $\rho_0^{\mathrm {II}}$ are initial densities for the two phases, respectively. Thereafter, $\rho_{\mathrm{ref}}^{\mathrm {I}}=0.99\rho_{min}\rho _{0}^{\mathrm {I}}$, $\rho_{\mathrm{ref}}^{\mathrm {II}}=0.99\rho_{min}\rho _{0}^{\mathrm {II}}$ are subtracted as reference density
for each phase in Eq.~(\ref{state}) to compute the particle pressure.
This operation is performed to reduce numerical errors
in calculating the pressure gradient
while still keeping repulsive forces between particles. 

The explicit velocity-Verlet method is adopted for time integration
and a time step is chosen appropriately for stability~\cite{morris2000SimulatingSurfaceTension}.

\section{Numerical Results}
\label{sec:results}

\begin{figure} 
\centering
\includegraphics[width=0.6\textwidth]{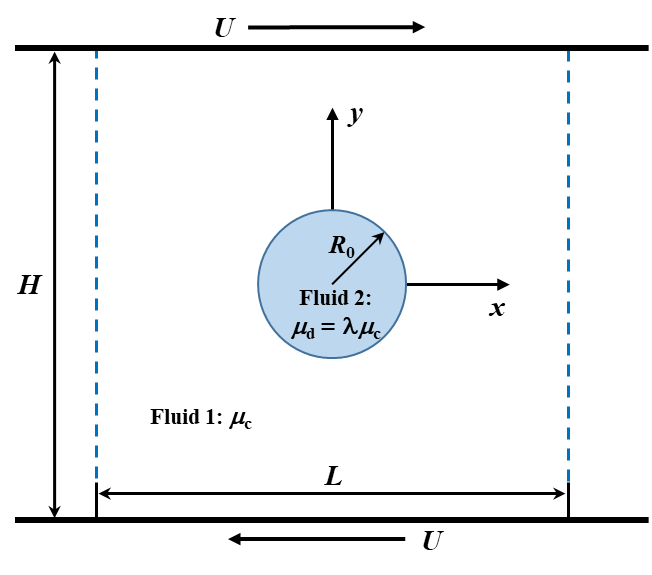} 
\caption{\label{fig:geometry}Schematic representation of a droplet with initial radius $R_{0}$ in a matrix fluid between two parallel walls with distance $H$. The blue dashed lines represent periodic boundaries with a distance $L$. The continuous phase has viscosity $\mu_c$ while the dispersed phase has viscosity $\mu_d=\lambda \mu_c$.}
\end{figure}

We consider a shear flow generated by two parallel walls with opposite velocity of magnitude $U$. Periodic boundaries apply in the $x$ direction.
The computational domain is with length $L$ and height $H$. A circular droplet with radius $R_{0}$ is initially located at the center of the computational domain, as shown in Fig.~\ref{fig:geometry}.
No-slip boundary condition is applied at the wall-fluid interfaces using the method proposed by Morris~\cite{morris1997ModelingLowReynolds}.

Five dimensionless parameters that determine the deformation of the droplet are Reynolds number $Re = \rho_c \dot{\gamma} R_{0}^{2} / \mu _{c}$, Capillary number $Ca = \dot{\gamma} R_{0}\mu _{c} / \sigma $, confinement ratio $R_{0}/H$,  viscosity ratio $\lambda = \mu_{d} / \mu _{c} $ and density ratio $\alpha = \rho_{d} / \rho _{c}$, where $\dot{\gamma} = 2U/H$ is the shear rate, $\sigma $ is the surface tension coefficient, $\rho_{d}$ and $\mu_{d}$ are density and viscosity of the dispersed fluid phase inside the droplet while and $\rho_{c}$ and $\mu _{c}$ are for the continuous fluid phase, respectively.

In Sec.~\ref{sec:result_dropdeform}, we study the deformation for an intact droplet while considering the effects due to the five dimensionless numbers. In Sec.~\ref{sec:result_breakup}, we examine the breakup of the droplet.
In Sec.~\ref{sec:result_phasediag}, we summarize the droplet dynamics for both intact shape and breakup in phase diagrams.
In Sec.~\ref{sec:result_waterair}, we demonstrate the deformation and breakup with physical parameters of a water droplet in air flow as an industrial application.

\subsection{Droplet deformation}
\label{sec:result_dropdeform}

\begin{figure} 
\centering
\includegraphics[width=0.8\textwidth]{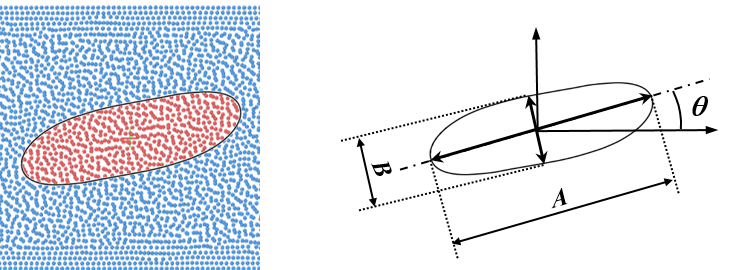} 
\caption{\label{fig:interface_taylor}Trace of interface and parameters for the measurement of droplet deformation.}
\end{figure}
When the shear is mild, the droplet remains intact and deforms to arrive at a stable shape eventually.
The degree of droplet deformation can be quantified by the Taylor deformation parameter $D = (A-B)/(A+B)$, where $A$ is the greatest length and $B$ is the breadth of the droplet as shown in Fig.~\ref{fig:interface_taylor}. To validate our method, we first compare our results of transient deformations with that of Sheth and Pozrikidis using immersed boundary method within the finite difference method~\cite{sheth1995EffectsInertiaDeformation}.
We follow their work to set $L=H=4R_{0}=1$, $\rho_d = \rho_c = 1$, $\mu_{d} = \mu _{c} = 0.5$ and adjust shear rate and surface tension.
The two walls slide with velocities $\pm \frac{1}{2} \dot{\gamma}H$
to generate a clockwise rotation of the droplet.
Two resolutions are considered for particles initially placed on squared lattice: $\Delta x = 2R_0/25$ and $R_0/25$, corresponding to the droplet containing $N=484$ and $1976$ particles, respectively.

\begin{figure} 
\centering
\subfigure[$\dot{\gamma}t = 0$]
{
\centering
\includegraphics[scale=0.21]{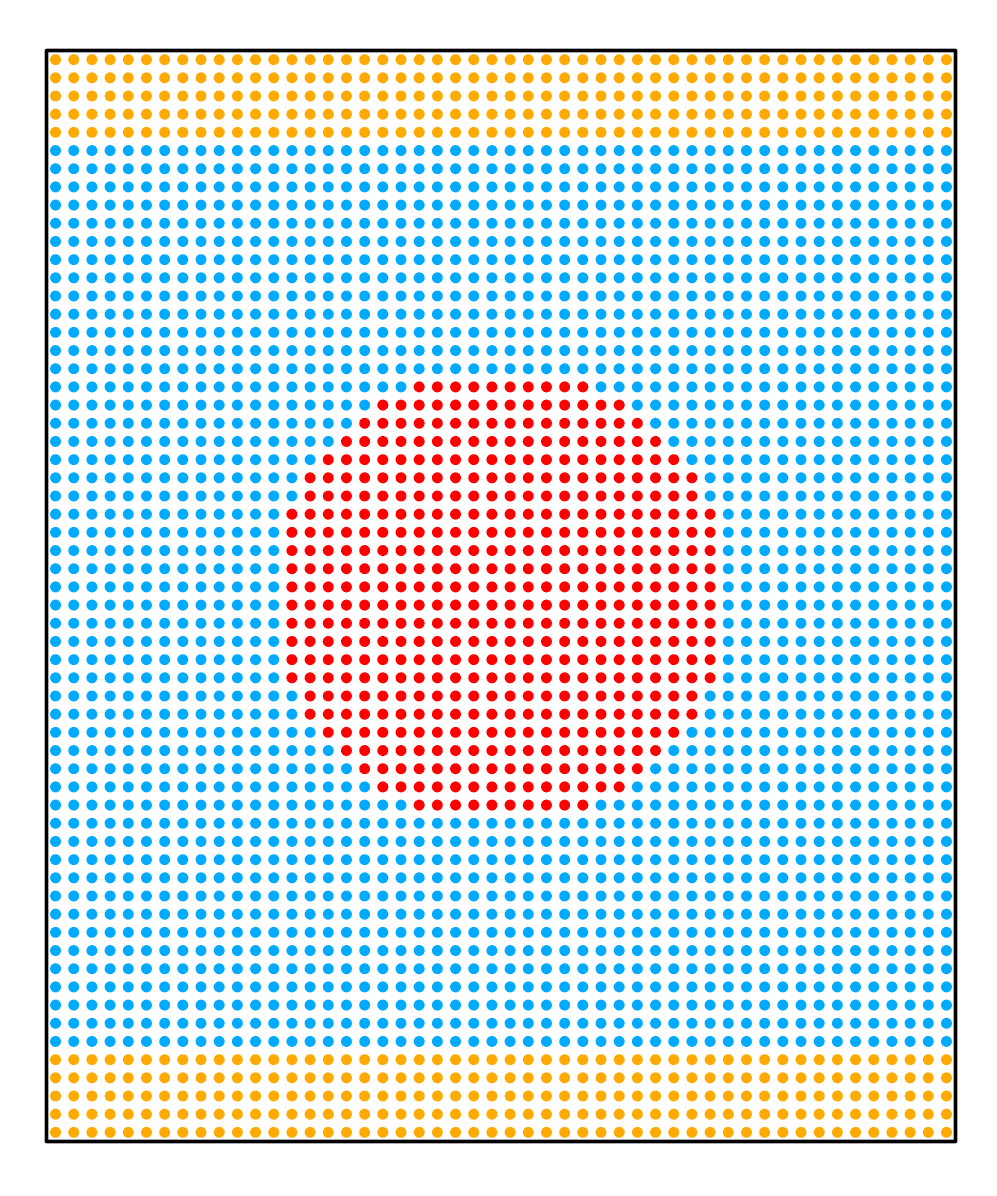} 
}
\subfigure[$\dot{\gamma}t = 4$]
{
\centering
\includegraphics[scale=0.21]{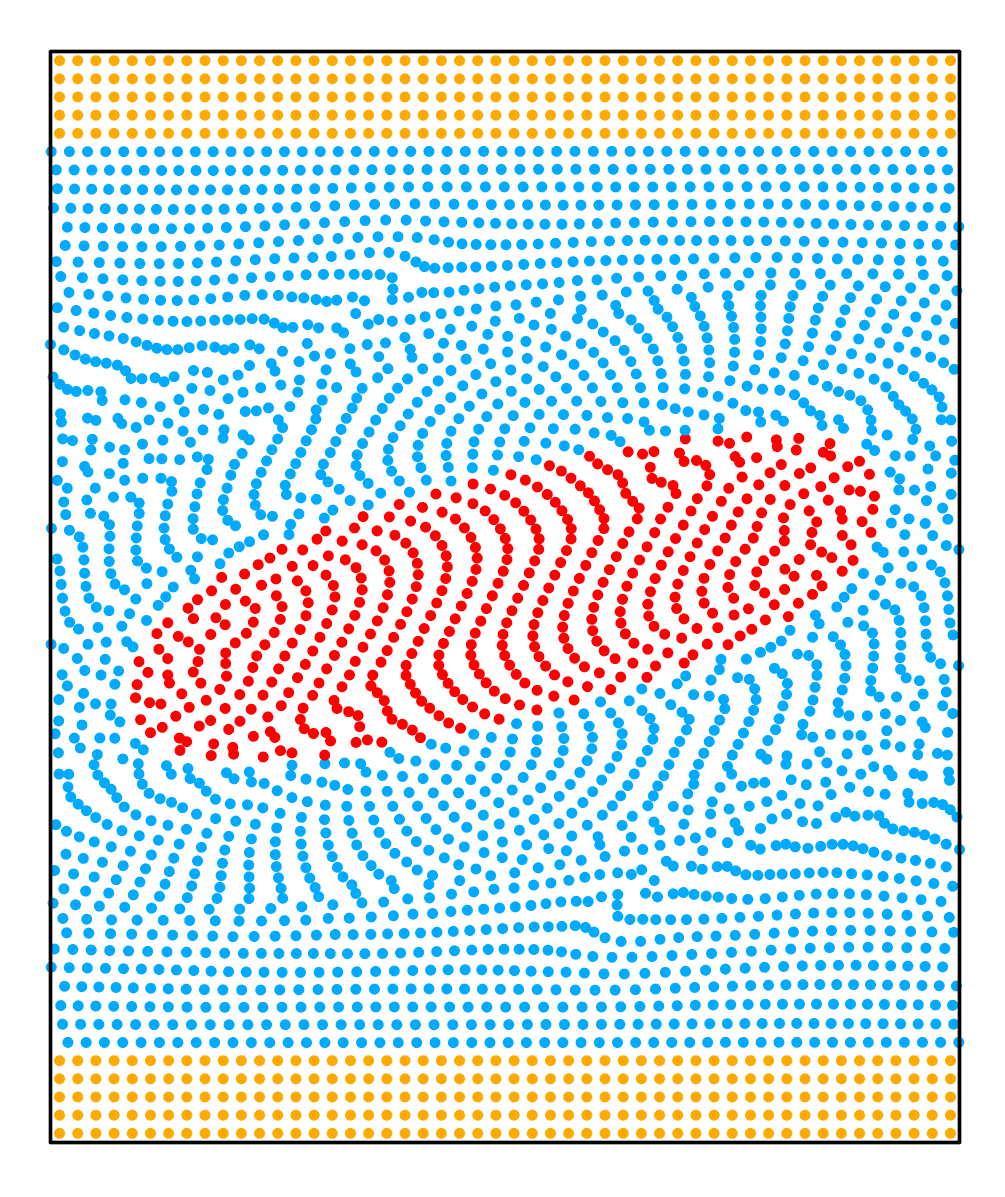} 
}
\subfigure[$\dot{\gamma}t = 20$]
{
\centering
\includegraphics[scale=0.21]{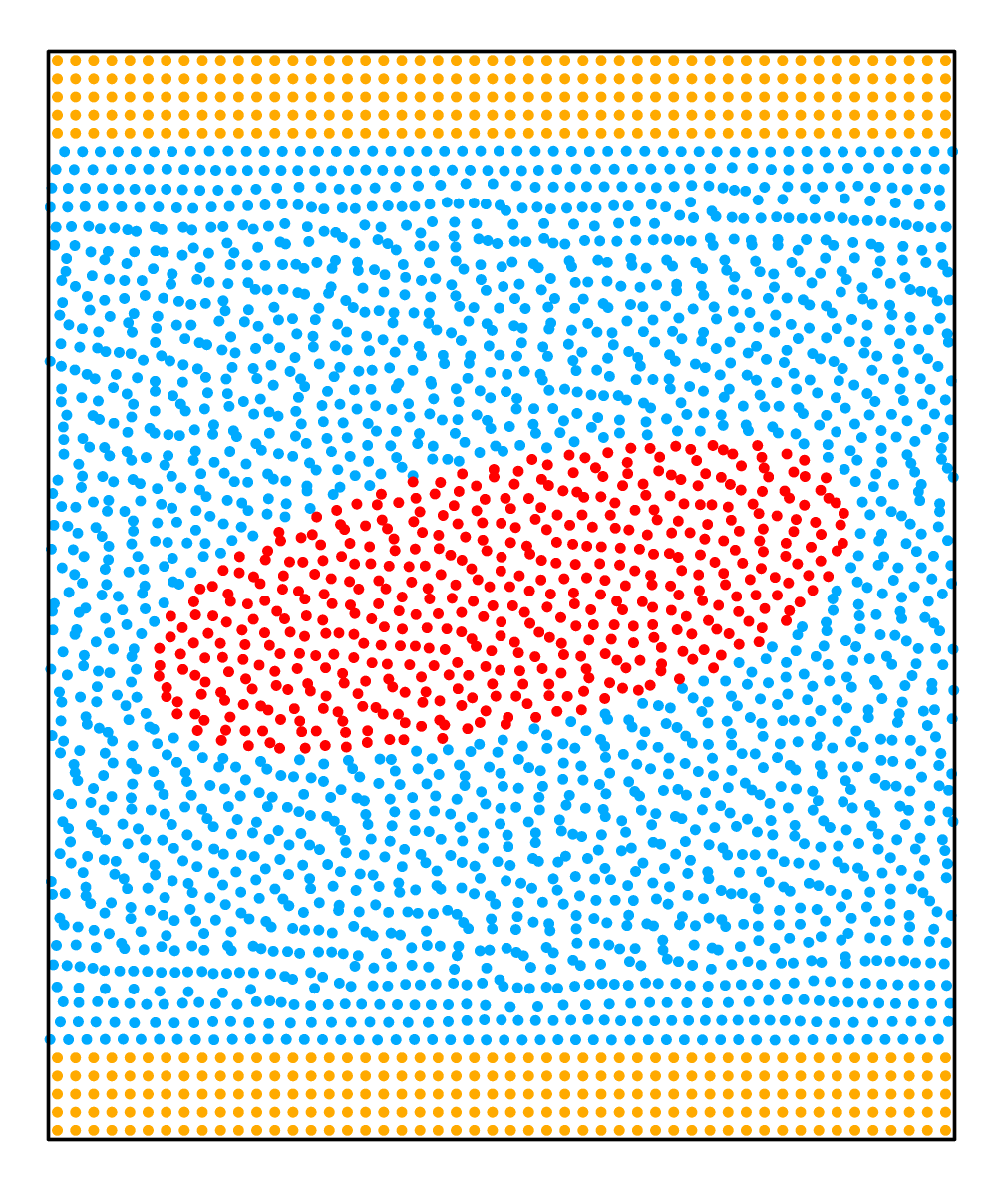} 
}
\subfigure[]
{
\centering
\includegraphics[scale=0.56]{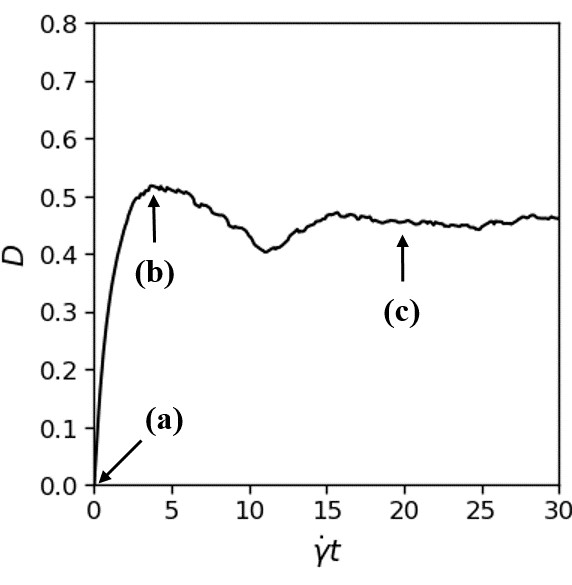} 
}
\caption{\label{fig:sheth1995particle}{Particle distribution for  $H=4R_0$, $\alpha=\lambda=1$, $Re=0.125$, $Ca=0.45$, $L=4R_0$ at (a) initial configuration. (b) maximum elongation. (c) steady state. (d) Taylor deformation parameter as a function of time.}
}
\end{figure}
We present particle distributions and $D$ as functions of time in Fig.~\ref{fig:sheth1995particle} for a typical simulation with $Re=0.125$ and $Ca=0.45$.
We note that the deformation of the droplet may oscillate in time
and its maximum elongation does not necessarily takes place at the steady state of a very long time.
\begin{figure*} 
\centering
\subfigure[$Re = 0.125$]
{
\centering
\includegraphics[scale=0.5]{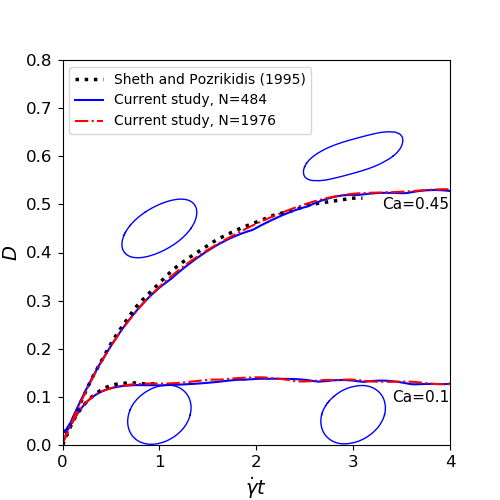} 
}
\subfigure[$Re = 1.25$]
{
\centering
\includegraphics[scale=0.5]{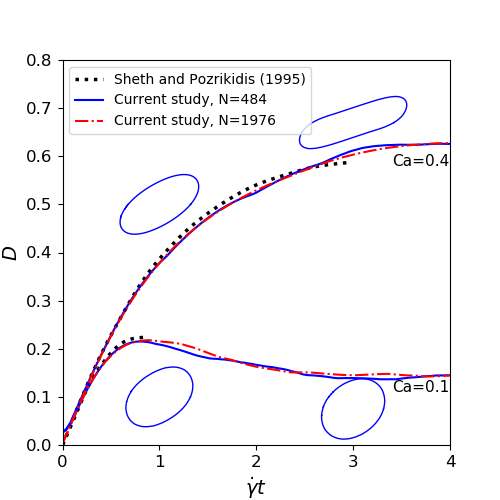} 
}

\subfigure[$Re = 6.25$]
{
\centering
\includegraphics[scale=0.5]{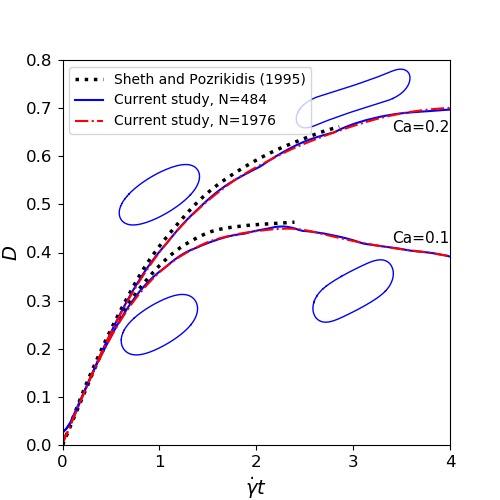} 
}
\subfigure[$Re = 12.5$]
{
\centering
\includegraphics[scale=0.5]{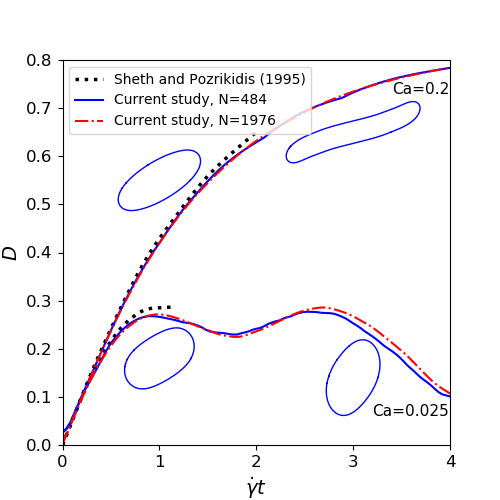} 
\label{fig:comp_sheth1995_xall_d}
}
\caption{\label{fig:comp_sheth1995_xall}Transient deformations of a droplet compared with results of Sheth and Pozrikidis~\cite{sheth1995EffectsInertiaDeformation} with  $\alpha=\lambda=1$, $H=4R_0$, $L=4R_0$ and (a) $Re=0.25$, $Ca=0.1,\,0.45$. (b) $Re=2.5$, $Ca=0.1,\,0.4$. (c) $Re=12.5$, $Ca=0.1,\,0.2$. (d) $Re=25$, $Ca=0.025,\,0.2$}
\end{figure*}
We further focus on the transient deformations in short time in Fig.~\ref{fig:comp_sheth1995_xall} so that we can compare our results with those of Sheth and Pozrikidis~\cite{sheth1995EffectsInertiaDeformation}.
It can be readily seen that 
our results with low resolution $\Delta x = 0.02$ or $N=484$ already reproduce the reference very well for different Reynolds numbers and/or capillary numbers.
As the reference is within a rather short time period,
some interesting phenomenon such as oscillation of the Taylor deformation parameter $D$ is not captured, as indicated for $Re=12.5$ and $Ca=0.025$ on Fig.~\ref{fig:comp_sheth1995_xall_d}.

\begin{figure*} 
\centering
\begin{minipage}[legend]{0.49\linewidth} 
    \subfigure[]
    {
    \label{zhou1993_cad}\includegraphics[scale=0.5]{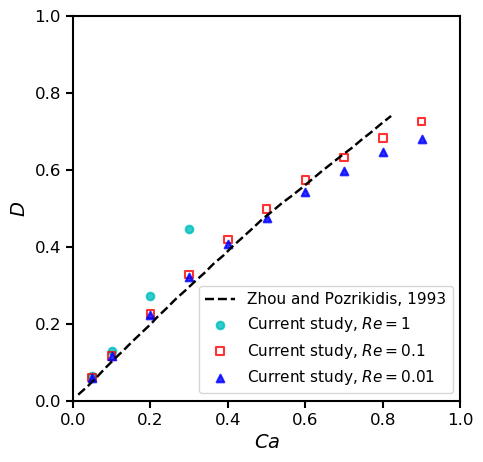} 
    }
    \subfigure[]
    {
    \label{zhou1993_cat}\includegraphics[scale=0.5]{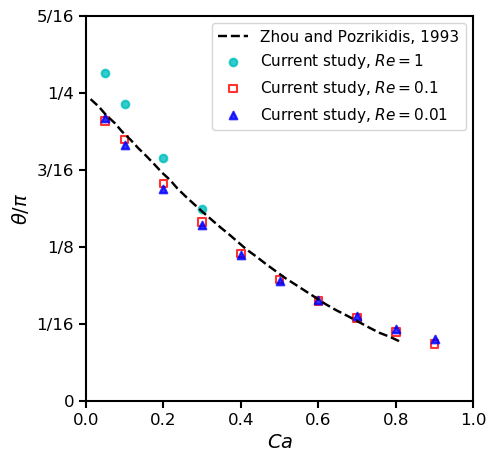} 
    }
\end{minipage}
\begin{minipage}[legend]{0.49\linewidth} 
    \subfigure[]
    {
    \label{zhou1993_shape}\includegraphics[scale=0.64]{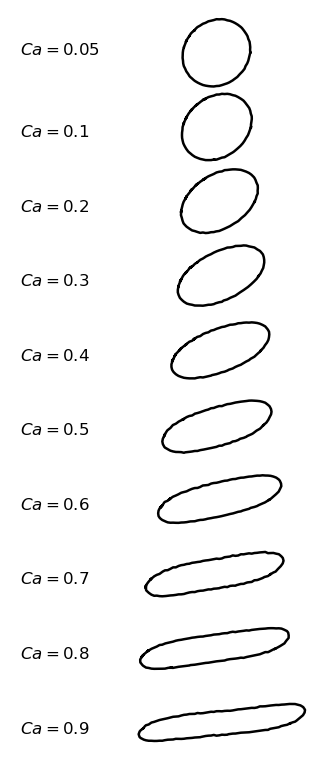} 
    }
\end{minipage}
\caption{\label{fig:comp_zhou1993}Effects of $Re$ and $Ca$ on 2D droplet deformation when $\alpha=\lambda=1$, $H=8R_0$, $L=8R_0$. (a) Taylor deformation parameters. (b) Droplet orientation. (c) Steady shapes under different $Ca$ when $Re=0.1$. The droplet already breaks up at $Ca \gtrsim 0.4$ for $Re=1$, results of which are omitted here and presented in a later section.}
\end{figure*}
To validate our method for vanishing Reynolds numbers, we calculate the stationary deformation and orientation of the droplet with respect to $Ca$. We follow Zhou and Pozrikidis~\cite{zhou1993FlowSuspensionsChannels} to set $L=H=8R_{0}=2$, $\rho_d = \rho_c =1$, $\mu_{d} = \mu _{c} = 0.5$ and adjust shear rate and surface tension accordingly.  The deformation parameter $D$ and orientation $\theta$~(defined on Fig.~\ref{fig:interface_taylor}) as functions of $Ca$ (up to $Ca=1$) for $Re=0.01$ are shown in Fig.~\ref{fig:comp_zhou1993}. Results for $Re=0.1$ and $1$ are also given for comparison,
where droplet breakup already takes places at $Ca \gtrsim 0.4$ for $Re=1$. The difference between the results of $Re=0.1$ and $Re=0.01$ is insignificant and they both resemble the results of boundary integral method for Stokes flow~\cite{zhou1993FlowSuspensionsChannels}. We can readily conclude that $Re=0.1$ is small enough to approximate the Stokes flow
and present the steady shapes accordingly on Fig.~\ref{zhou1993_shape}.
\begin{figure*} 
\centering
\includegraphics[width=0.88\textwidth]{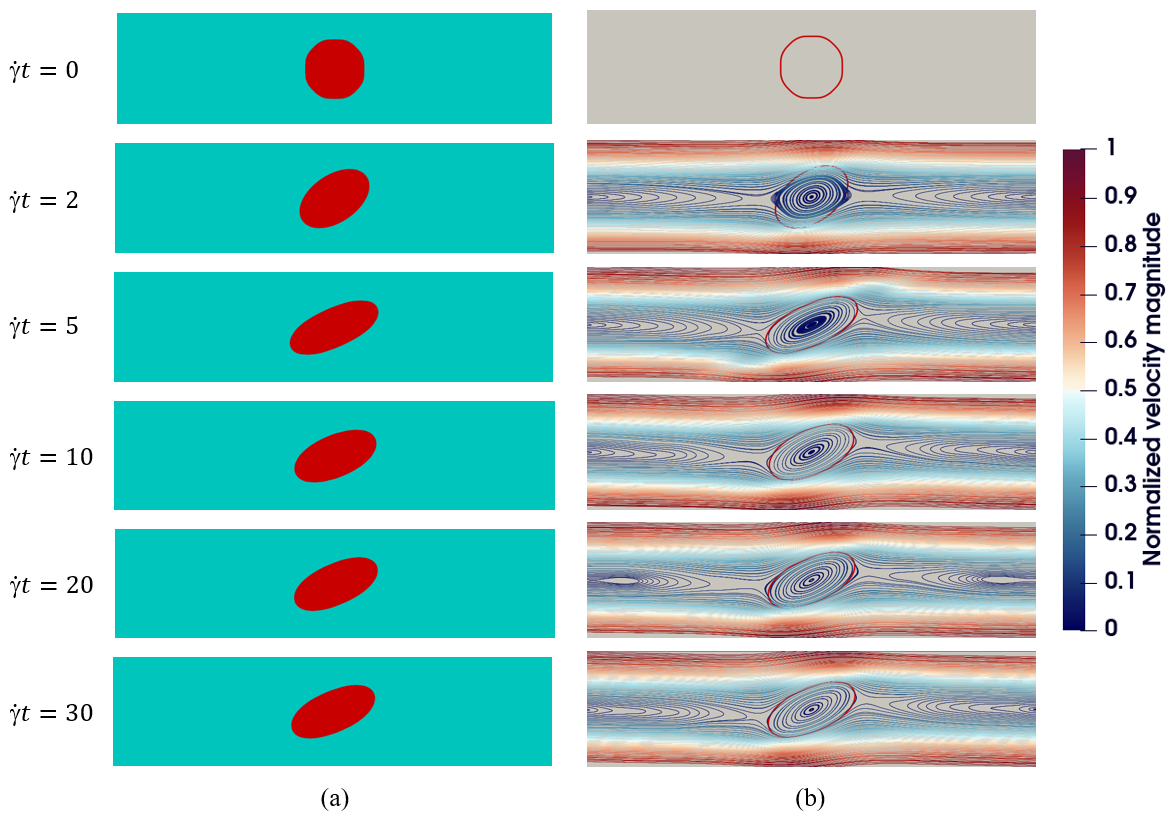}
\caption{\label{fig:stream_steady2} A typical evolution of deformation for an initially circular 2D droplet in shear flow: $\alpha=\lambda=1$, $Re=0.1$, $Ca=0.4$, $H=16R_0$, $L=16R_0$. (a) Droplet deformation over time. (b) Streamlines: the color represents the magnitude of velocity and a red line indicates the droplet interface.}
\end{figure*}
We further present the contours and streamlines for a typical evolution of droplet deformation at vanishing Reynolds number in Fig.~\ref{fig:stream_steady2}.

\begin{figure*} 
\centering
\subfigure[]
{
\centering
\includegraphics[scale=0.5]{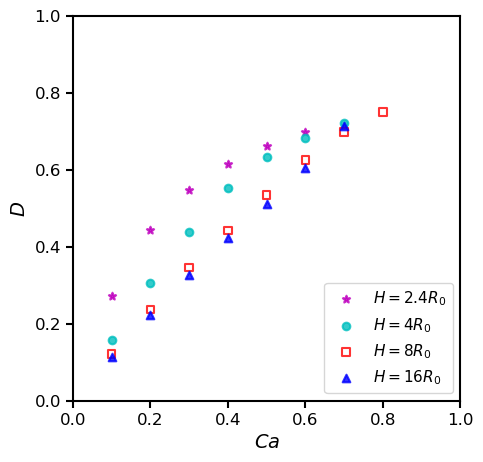} 
}
\subfigure[]
{
\centering
\includegraphics[scale=0.5]{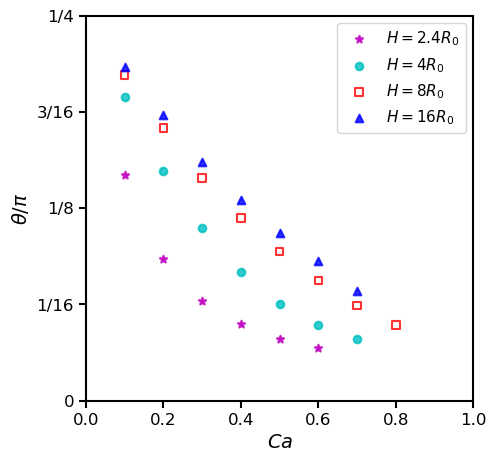} 
}
\caption{\label{fig:steay_conf}Effects of confinement ratio and $Ca$ on 2D droplet deformation when $\alpha=\lambda=1$, $Re=0.1$, $L=16R_0$. (a) Taylor deformation parameters. (b) Droplet orientation.}
\end{figure*}
We commence to investigate the effects of confinement and
set $L=16R_0$ to minimize the periodic artifacts.
We first restrict out attention to $Re=0.1$, $\alpha=1$ and $\lambda=1$.  
Four ratios of confinement are considered: $H=2.4R_0$, $4R_0$, $8R_0$ and $16R_0$.
The deformation parameter as a function of $Ca$ is shown in Fig.~\ref{fig:steay_conf}. As we can see, a smaller distance of the two walls enhances the elongation of droplet and makes its long axis align more horizontally. As we relax the confinement, the relation between $D$ and $Ca$ becomes linear and the difference between $H=16R_0$ and $H=8R_0$ is already negligible.

\begin{figure*} 
\centering
\subfigure[]
{
\centering
\includegraphics[scale=0.5]{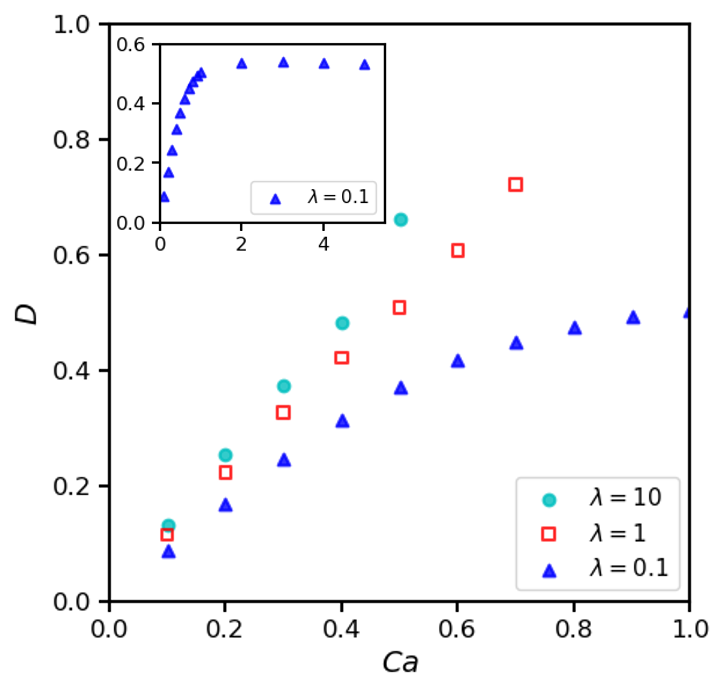} 
}
\subfigure[]
{
\centering
\includegraphics[scale=0.5]{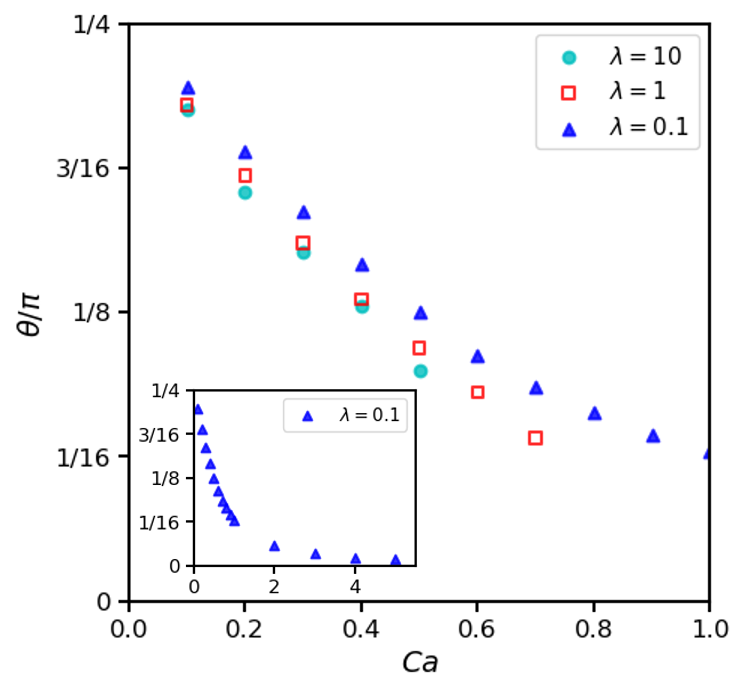} 
}
\caption{\label{fig:ca_lam_d}Effects of viscosity ratio $\lambda$ and $Ca$ on 2D droplet deformation when $\alpha=1$, $Re=0.1$, $H=16R_0$, $L=16R_0$. (a) Taylor deformation parameters. (b) Droplet inclination.}
\end{figure*}
Furthermore, we simulate cases where the droplet and the matrix flow are two fluids with different physical properties. We first consider two fluids of the same density but with different viscosities. We choose a computational domain of $16R_0 \times 16R_0$ and set $Re=0.1$, $\alpha=1$ and $\lambda$ ranges from 0.1 to 10. Initial space $\Delta x$ among nearest particles is $2R_0/25$ so a droplet contains 484 particles. The deformation parameter as a function of $Ca$ is shown in Fig.~\ref{fig:ca_lam_d}. As we can see, the deformation increases as $\lambda$ increases from $0.1$ to $10$. In this range of $\lambda$, a droplet with lower viscosity has a smooth inside circulation and fast reaction which can reduce the elongation~\cite{karam1968deformation, grace1982dispersion}.

\begin{figure*} 
\centering
\subfigure[]
{
\centering
\includegraphics[scale=0.5]{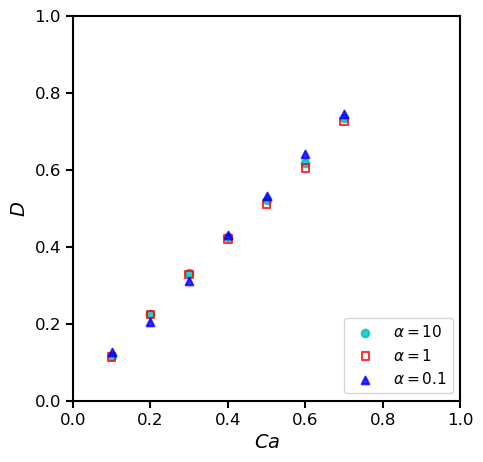} 
}
\subfigure[]
{
\centering
\includegraphics[scale=0.5]{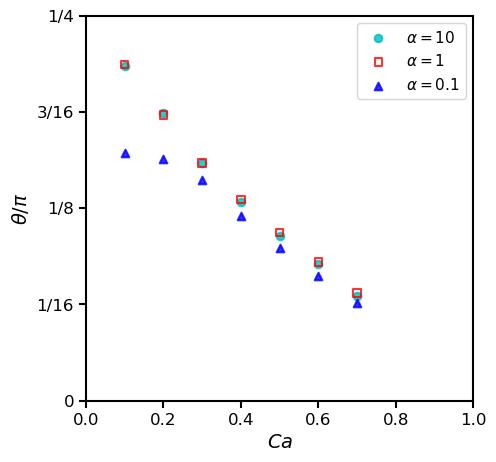} 
}
\caption{\label{fig:ca_rho_d}Effects of density ratio $\alpha$ and $Ca$ on 2D droplet deformation when $\lambda=1$, $Re=0.1$, $H=16R_0$, $L=16R_0$. (a) Taylor deformation parameters. (b) Droplet inclination.}
\end{figure*}
The other case is that fluids inside and outside the droplet have the same viscosity but different densities. The sound speed is chosen according to the ratio of initial density to  balance the pressure
\begin{equation}
\frac{c_{s}^{c}}{c_{s}^{d}} = \frac{\rho_{ref}^{d}}{\rho_{ref}^{c}} = \frac{\rho_{d}}{\rho_{c}}=\alpha,
\label{cs_diff}
\end{equation}
where $c_{s}^{c}$, ${c_{s}^{d}}$ and $\rho_{ref}^{c}$, $\rho_{ref}^{d}$ are sound speeds and reference densities used for fluids outside and inside the droplet. As shown in Fig.~\ref{fig:ca_rho_d}, the difference between deformations of droplet under density ratio $0.1 - 10$ is very small except obvious lower inclination at small $Ca$ when $\alpha=0.1$. In this small Reynolds number regime~($Re=0.1$), the density ratio has negligible influence and only the capillary number determines the droplet deformation.

\begin{figure*} 
\centering
\subfigure[]
{
\centering
\includegraphics[scale=0.6]{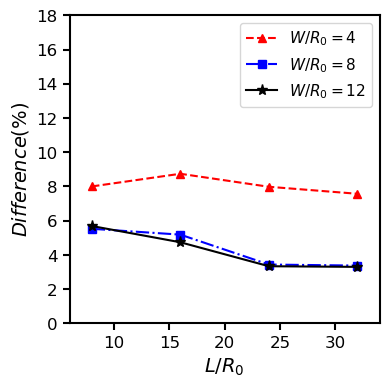} 
}
\subfigure[]
{
\centering
\includegraphics[scale=0.6]{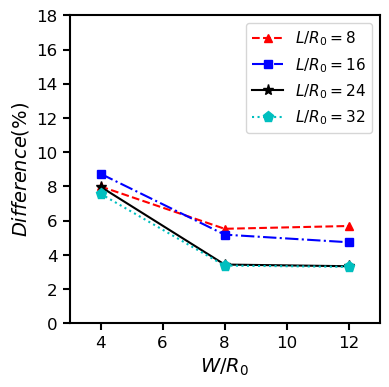} 
}
\caption{\label{fig:width_length}Effects of box length $L$ and width $W$ on
the Talor deformation parameter $D$ of 3D droplet deformation when $\alpha=\lambda=1$, $Re=0.1$, $Ca=0.2$, $H=4R_0$, compared to the analytical prediction of Shapira and Haber (1990)}
\end{figure*}

\begin{figure} 
\centering
\includegraphics[width=0.55\textwidth]{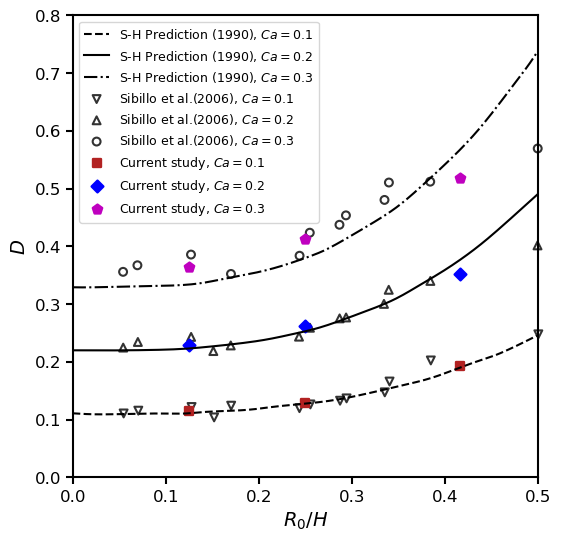} 
\caption{\label{fig:steady_3d_conf} Effects of confinement ratio and $Ca$ on 3D droplet deformation in shear flow when $\alpha=1$ and $Re=0.1$, $L=24R_0$, $W=8R_0$, compared with predictions of Shapira and Haber~\cite{shapira1990LowReynoldsNumber} and experiment data of Sibillo et al.~\cite{sibillo2006DropDeformationMicroconfined}}.
\end{figure}
In 3D simulations, the width of simulation box $W$ is an additional computational parameter compared to the 2D simulations. To compare with analytical predictions or experiment data, the length and width of simulation box are numerical and should be large enough. One set of parameters of $Re=0.1$, $Ca=0.2$, $\alpha = \lambda =1$, $H = 4R_0$ are selected and different length $L$ and width $W$ of simulation box are examined. According to our simulations, the deformation basically decreases with the increase of $L$ and/or width $W$. We compare the Taylor deformation parameter $D$ in steady states of our simulations with the analytical prediction of Shapira and Haber~\cite{shapira1990LowReynoldsNumber}. The differences between our results and analytical prediction under different $L$ and $W$ are plotted in Fig.~\ref{fig:width_length}. It can be seen that when $L$ is larger than $24R_0$ and $W$ is larger than $8R_0$, the results has little change with the increase of $L$ and/or $W$. Fig.~\ref{fig:steady_3d_conf} shows the steady deformation of 3D droplets in shear flow when $L=24R_0$, $W=8R_0$, $Re=0.1$ and $\alpha=\lambda =1$ with different $Ca$ and confinement in $H$ direction, compared with theoretical predictions of Shapira and Haber~\cite{shapira1990LowReynoldsNumber} and experiment data of Sibillo et al.~\cite{sibillo2006DropDeformationMicroconfined}. Our results agree well with both anlaytical and experiment references at $Ca=0.1$ and $0.2$, 
whereas are closer to the experimental data at $Ca=0.3$. The deformation increases with the confinement ratio $R_0/H$, which has the same trend as for 2D cases.

\subsection{Droplet breakup}
\label{sec:result_breakup}

\begin{figure*} 
\centering
\includegraphics[width=0.88\textwidth]{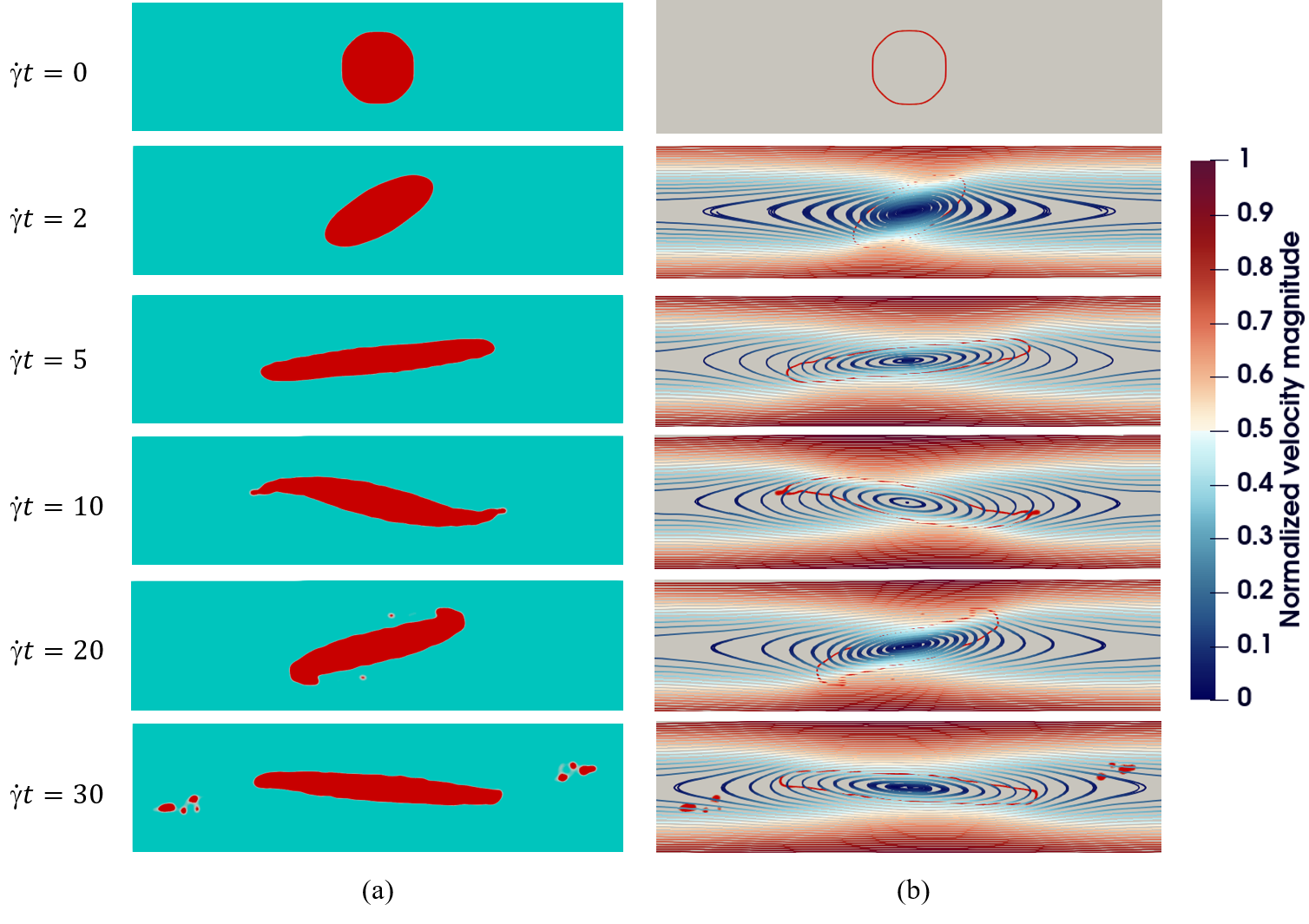}
\caption{\label{fig:stream_break_tear} Breakup type A: evolution of an initially circular 2D droplet breakup in shear flow with $\alpha=1$, $\lambda=0.2$, $Re=0.1$, $Ca=10$, $H=16R_0$, $L=16R_0$. (a) Droplet shape. (b) Streamlines: The color represents the magnitude of velocity and red outlines in the background represent the droplet interface.}
\end{figure*}
\begin{figure*} 
\centering
\includegraphics[width=0.88\textwidth]{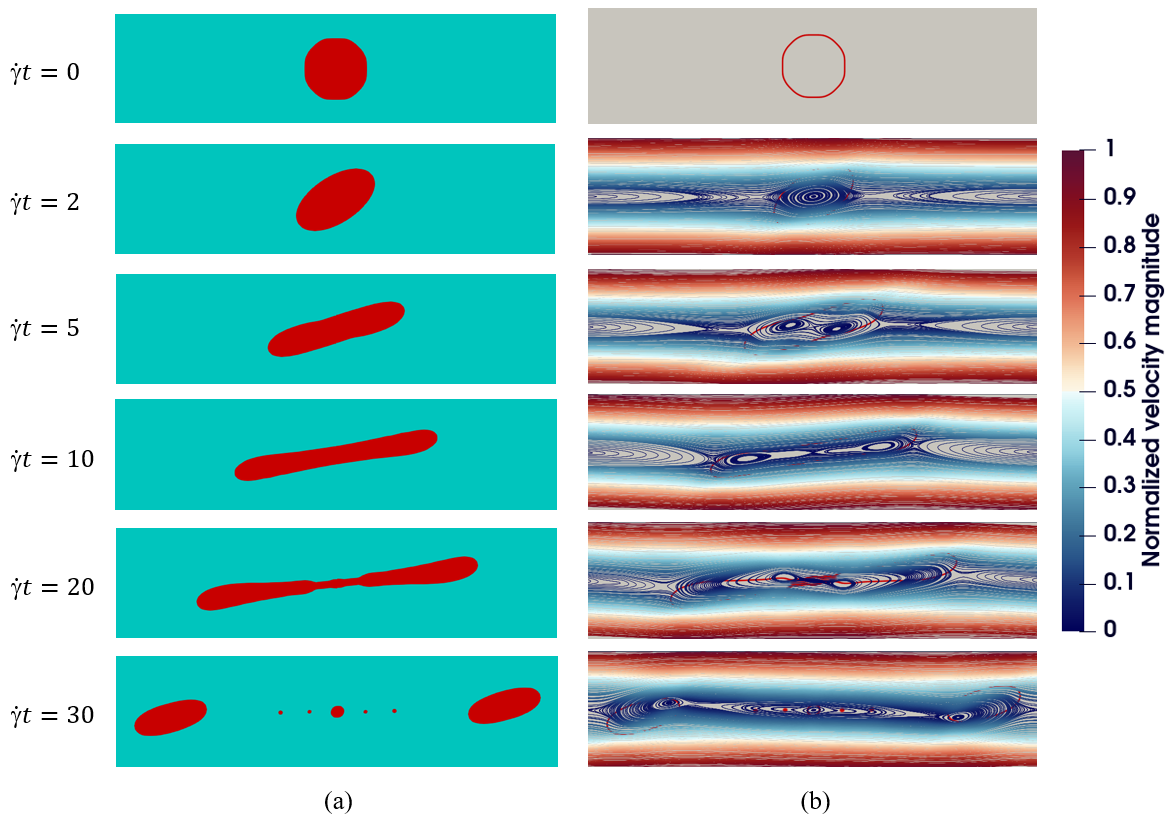}
\caption{\label{fig:stream_break2} Breakup type B: evolution of an initially circular 2D droplet breakup in shear flow with $\alpha=\lambda=1$, $Re=0.1$, $Ca=0.9$, $H=16R_0$, $L=16R_0$. (a) Droplet shape. (b) Streamlines: The color represents the magnitude of velocity and red outlines in the background represent the droplet interface.}
\end{figure*}
When the shear is strong, the droplet is over-stretched to break up. We find two patterns of breakup process under different viscosity ratios in simulations. As shown in Fig.~\ref{fig:stream_break_tear}, when $\alpha=1$, $\lambda=0.2$, $Re=0.1$, $Ca=10$, and $L=H=16R_0$, a droplet is rotated and then stripped of its main body near the surface and gradually breaks apart. We call this breakup type A. This type is also found in the experiment study of Grace and they call it "tip streaming breakup"~\cite{grace1982dispersion}. The conditions for type A breakup happening is exhibited in the next section. Fig.~\ref{fig:stream_break2} shows another set of typical snapshots of the droplet shape and flow fields in shear flow breaks when $\alpha=\lambda=1$, $Re=0.1$, $Ca=0.9$ and $L=H=16R_0$. In this simulation, a droplet is stretched and its waist becomes slender and slender and finally breaks up. We call this breakup type B.

\begin{figure*} 
\centering
\includegraphics[width=0.85\textwidth]{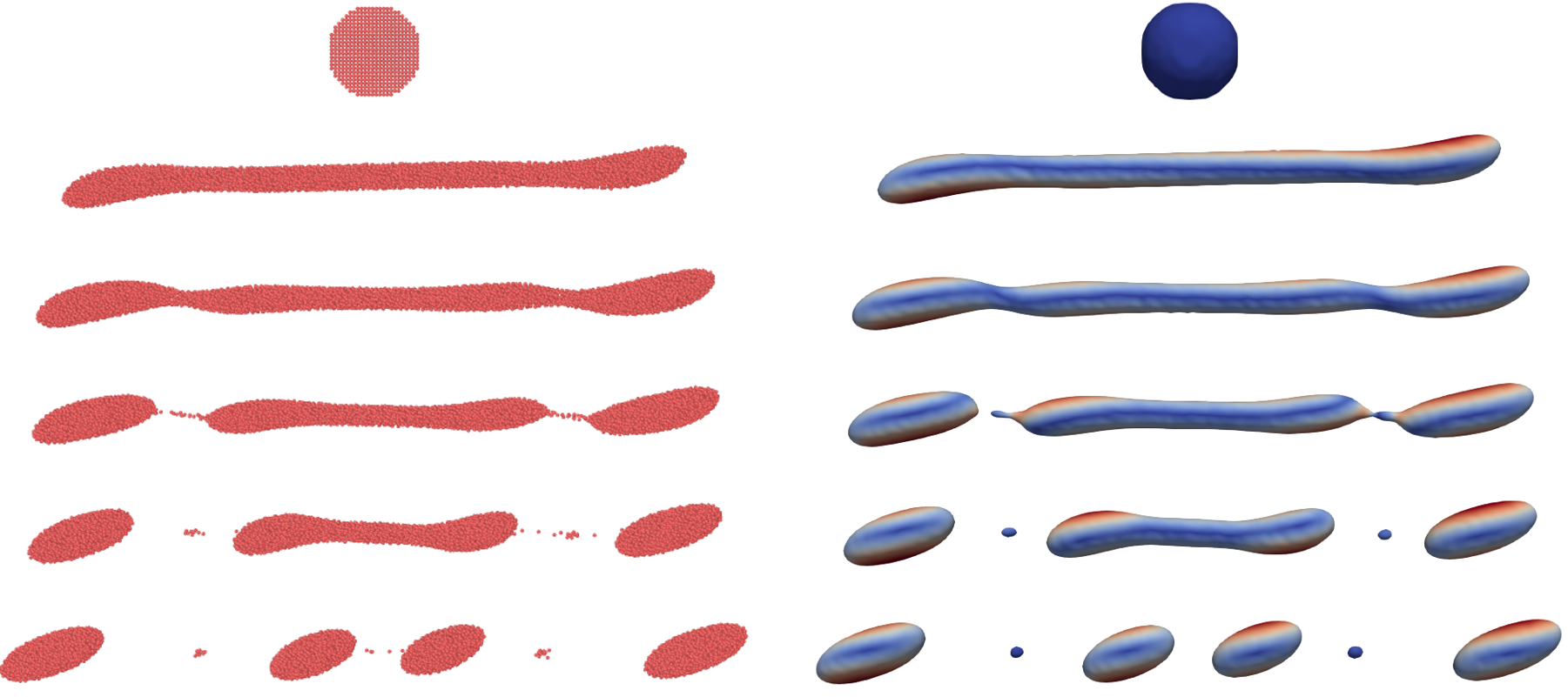}
\caption{\label{fig:break_3dsibilo} Evolution of an initially spherical 3D droplet breakup in shear flow when $\alpha=\lambda=1$, $Re=0.1$, $Ca=0.46$, $H=2.857R_0$, $L=16R_0$: particle distribution (left) and interface (right).}
\end{figure*}
To encompass the breakup of a 3D droplet with a large elongation, we employ a rather long computational domain with $L=32R_0$. Fig.~\ref{fig:break_3dsibilo} shows the dynamics of the breakup with $Re=0.1$, $H=2.857R_0$, $Ca=0.46$ and $\alpha=\lambda=1$. Left side are SPH particle distributions and right side are corresponding contour interfaces processed by SPH kernel interpolation into mesh cells. The color represents the magnitude of velocity. We adopt the same $Ca$ and $R_0/H$ as the experiment in creeping flow by Sibillo et al.~\cite{sibillo2006DropDeformationMicroconfined}. The shape of the droplet in the breakup process of our simulation is very close to their experimental observation. Only a slight difference appears in the final stage: in the experiment, the droplet is divided into three main parts, while in our simulation the middle part continues to split into two smaller droplets. In contrast to the 2D case, a 3D droplet has a more slender shape before breaking up.

\subsection{Phase diagram}
\label{sec:result_phasediag}

To clearly visualize the states of a droplet in different conditions, we consider a range of Reynolds numbers, capillary numbers, and confinements/density/viscosity ratios
and summarize our simulation results into phase diagrams.
Thereafter, we may estimate the critical capillary number $Ca_c$ 
that segments the intact and breakup states and further investigate how it is influenced by other dimensionless parameters.

\begin{figure} 
\centering
\includegraphics[width=0.6\textwidth]{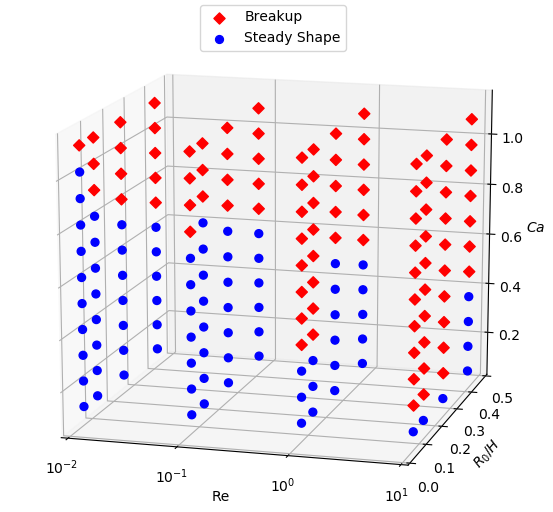} 
\caption{\label{fig:break_conf_re_all}Phase diagram of 2D droplets states under different confinement, $Re$ and $Ca$ when $\alpha=\lambda=1$, $L=16R_0$.}
\end{figure}

\begin{figure*} 
\centering
\subfigure[$H=2.4R_0$]
{
\centering
\includegraphics[scale=0.5]{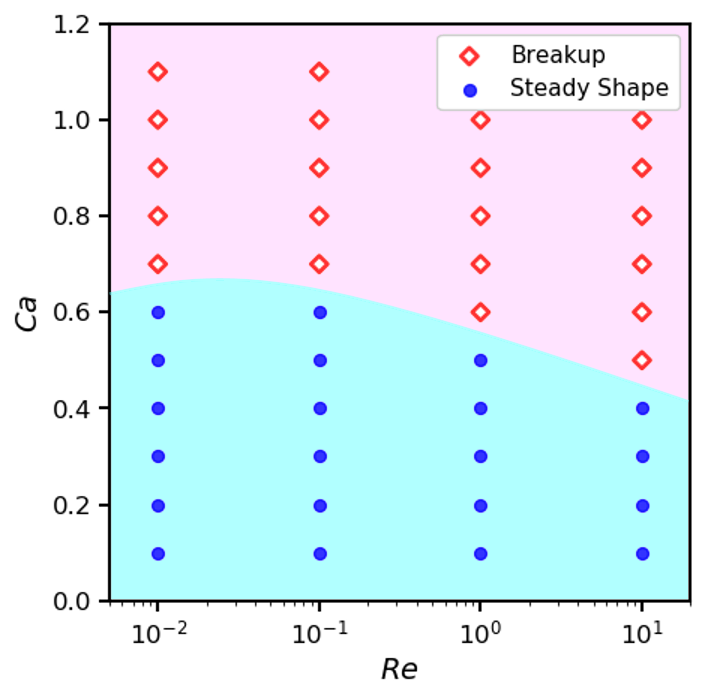} 
}
\subfigure[$H=4R_0$]
{
\centering
\includegraphics[scale=0.5]{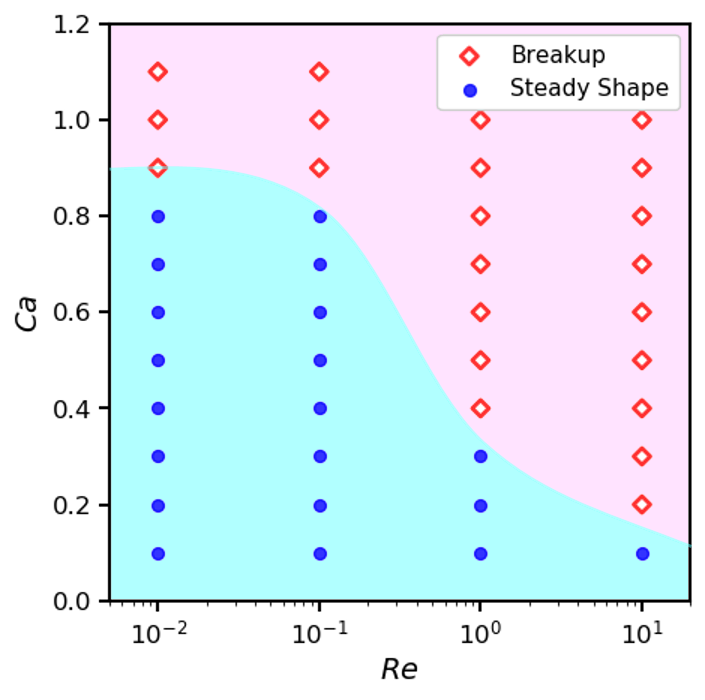} 
\label{fig:break_re_confb}
}
\subfigure[$H=8R_0$]
{
\centering
\includegraphics[scale=0.5]{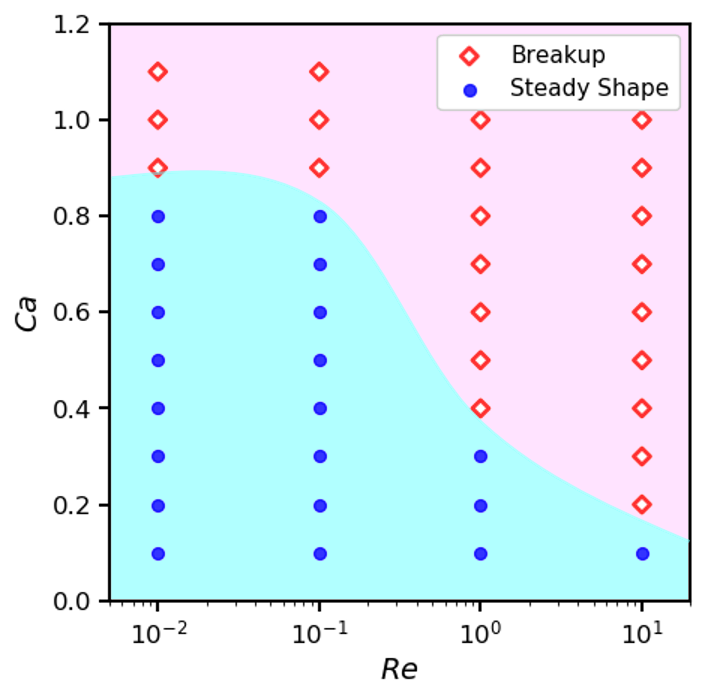} 
}
\subfigure[$H=16R_0$]
{
\centering
\includegraphics[scale=0.5]{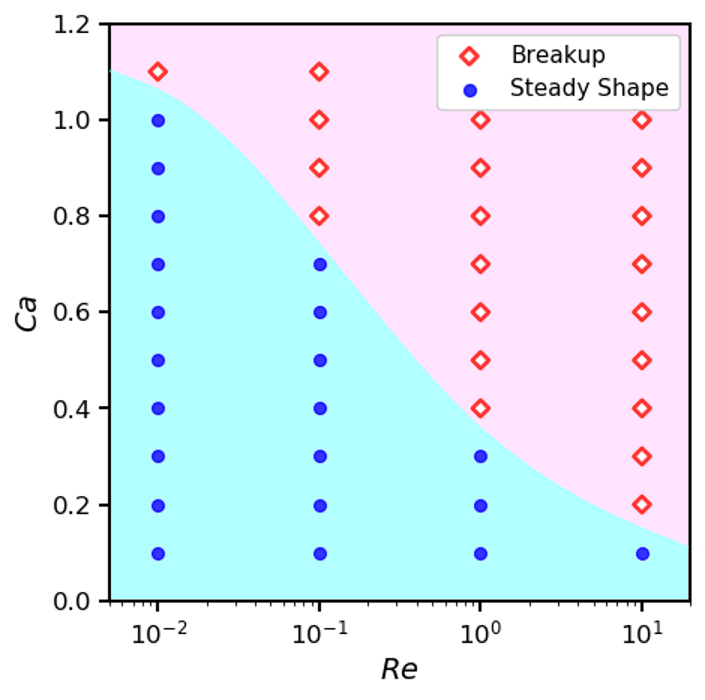} 
}
\caption{\label{fig:break_re_conf}Phase diagram of 2D droplets states under different confinement, $Re$ and $Ca$ when $\alpha=\lambda=1$, $L=16R_0$. (a) $H=2.4R_0$.(b) $H=4R_0$. (c) $H=8R_0$. (d) $H=16R_0$}
\end{figure*}
For $\lambda=\alpha=1$, we perform a group of 2D simulations with different Reynolds number $Re=0.01, 0.1,  1, 10$ and confinement $H=2.4R_0, 4R_0, 8R_0, 16R_0$ with $Ca \in [0.1, 1.1]$ and $L=16R_0$. 
For a general overview, the states of the droplet are summarized in Fig.~\ref{fig:break_conf_re_all}.
To get a clear view, we slice the phase diagram by two perspectives. Firstly, we divide results into groups of the same confinement to reveal the influence of $Re$ on $Ca_c$ as shown in Fig.~\ref{fig:break_re_conf}. 
Overall it is apparent that a higher $Re$ reduces $Ca_c$.
Three scenarios are special: under confinement $H=2.4R_0$, $4R_0$ and $8R_0$,
we can not differentiate $Ca_c$ between $Re=0.01$ and $0.1$.

\begin{figure*} 
\centering
\subfigure[$Re=0.01$]
{
\centering
\includegraphics[scale=0.5]{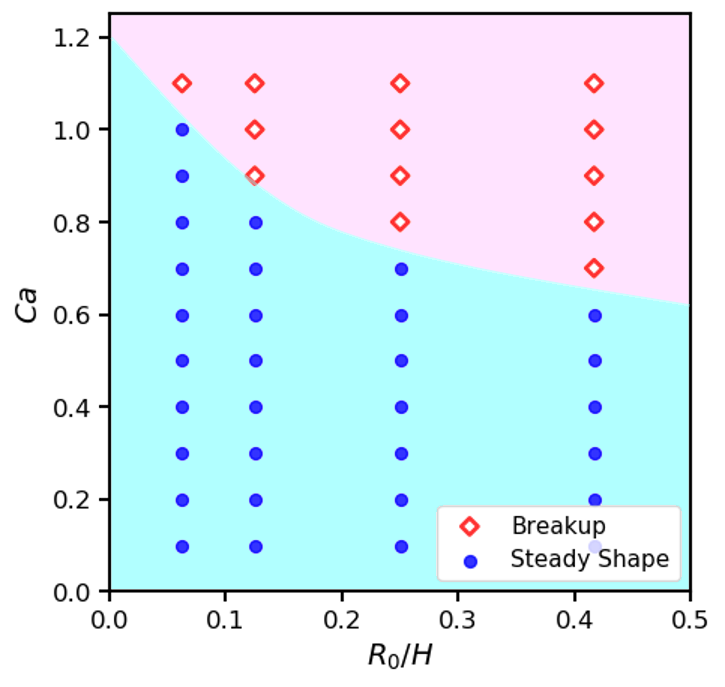} 
}
\subfigure[$Re=0.1$]
{
\centering
\includegraphics[scale=0.5]{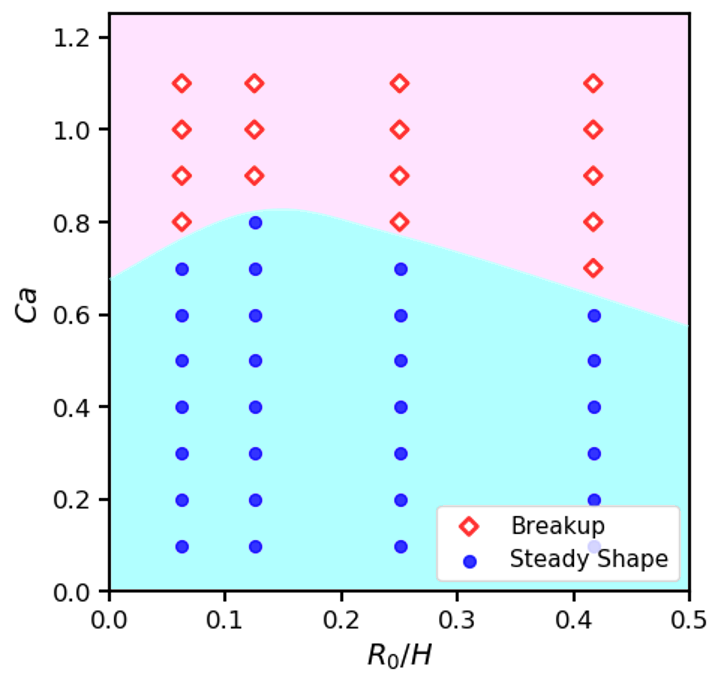} 
}
\subfigure[$Re=1$]
{
\centering
\includegraphics[scale=0.5]{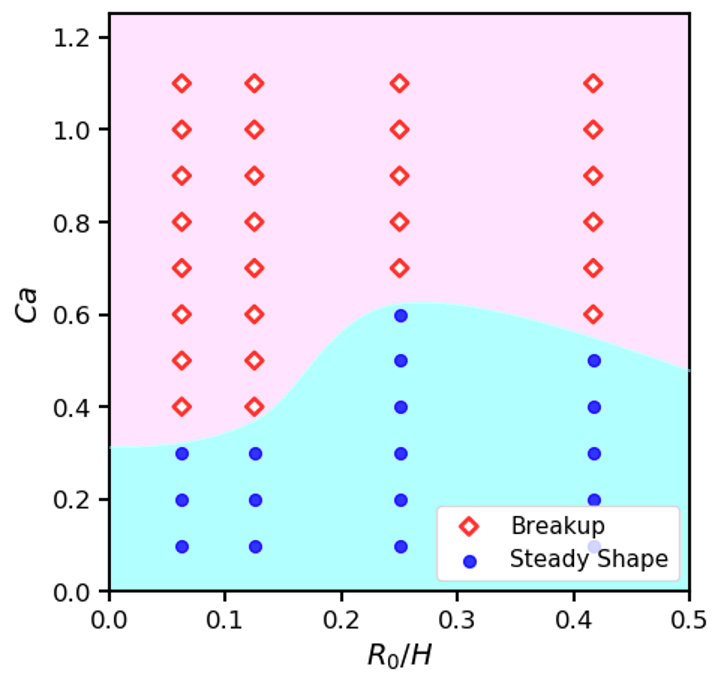} 
}
\subfigure[$Re=10$]
{
\centering
\includegraphics[scale=0.5]{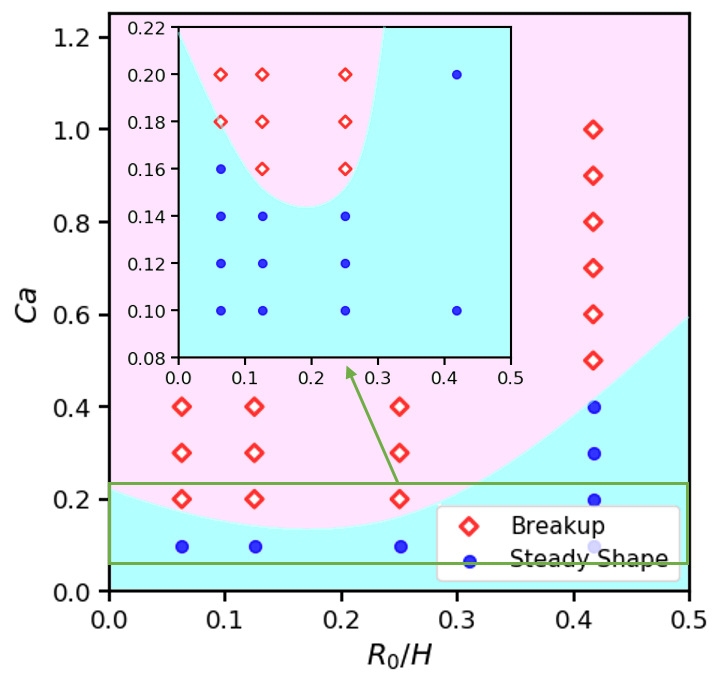} 
}
\caption{\label{fig:break_conf_re}Phase diagram of 2D droplets states under different confinement, $Re$ and $Ca$ when $\alpha=\lambda=1$, $L=16R_0$. (a) $Re=0.01$. (b) $Re=0.1$. (c) $Re=1$. (c) $Re=10$}
\end{figure*}

From another perspective of $Ca$ versus confinement ratio for each $Re$ on Fig.~\ref{fig:break_conf_re},
we are not able to find a universal pattern.
Under $Re=0.01$, $Ca_c$ decreases with $R_0/H$
while under $Re=10$, $Ca_c$ increases with $R_0/H$.
Whereas, under $Re=0.1$ and $1$, $Ca_c$ has no monotonic relation with $R_0/H$.

\begin{figure*} 
\centering
\subfigure[$H=4R_0$]
{
\centering
\includegraphics[scale=0.5]{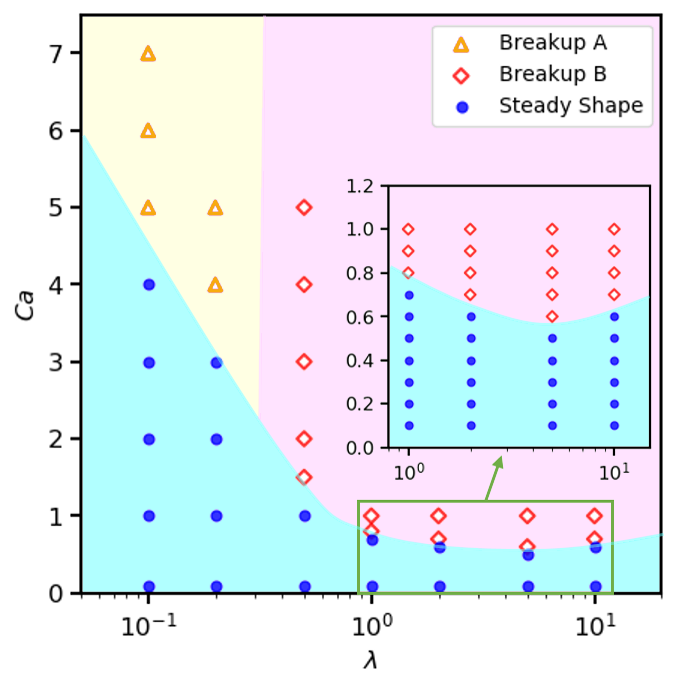} 
}
\subfigure[$H=8R_0$]
{
\includegraphics[scale=0.5]{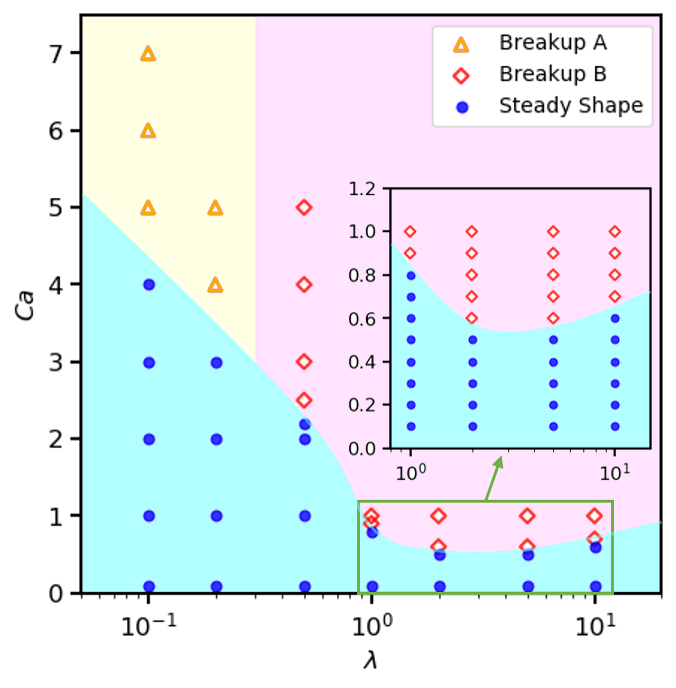} 
}
\subfigure[$H=16R_0$]
{
\centering
\includegraphics[scale=0.5]{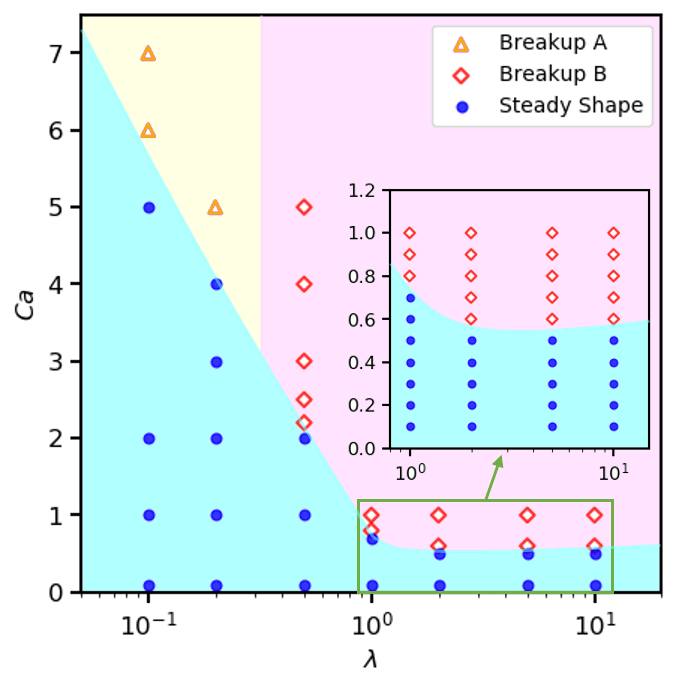} 
}
\subfigure[]
{
\centering
\includegraphics[scale=0.34]{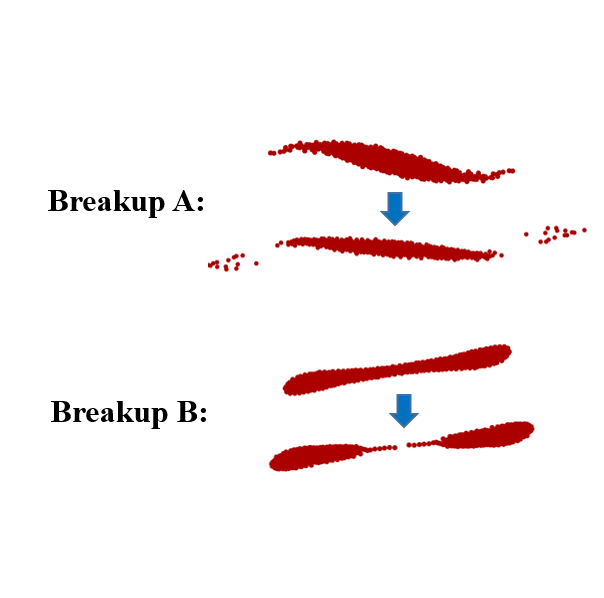} 
}
\caption{\label{fig:break_visco_con}Phase diagram of 2D droplets states under different confinement, viscosity ratios $\lambda$ and $Ca$ when $\alpha=1$, $Re=0.1$, $L=16R_0$. (a) $H=4R_0$. (b) $H=8R_0$. (c) $H=16R_0$. (d) two patterns of breakup}
\end{figure*}

Furthermore, we investigate effects of viscosity ratio $\lambda=\mu_d/\mu_c  \in [0.1, 10]$ on the droplet dynamics for $Re=0.1$ and three confinement ratios $H=4R_0$, $8R_0$ and $16R_0$. The results are shown in Fig.~\ref{fig:break_visco_con}. For breakup type A,  the droplet rotates and is stripped off as described in Sec.~\ref{sec:result_breakup};
Breakup type B represents that a droplet is stretched and breaks up in the middle.
Under $Re=0.1$, type A is observed only if the droplet has a much smaller viscosity compare to the matrix fluid.
Overall, $Ca_c$ decreases with the increase of $\lambda$.
However, we notice a flatten trend or even a reverse trend with small difference for $Ca_c$ from $\lambda=5$ to $\lambda=10$, as shown on the insets of Fig.~\ref{fig:break_visco_con}. 
According to the study of Karam et al. and Grace~\cite{karam1968deformation, grace1982dispersion}, a maximum transfer of energy takes place across an interface,
which demands this trend.

\begin{figure} 
\centering
\includegraphics[width=0.6\textwidth]{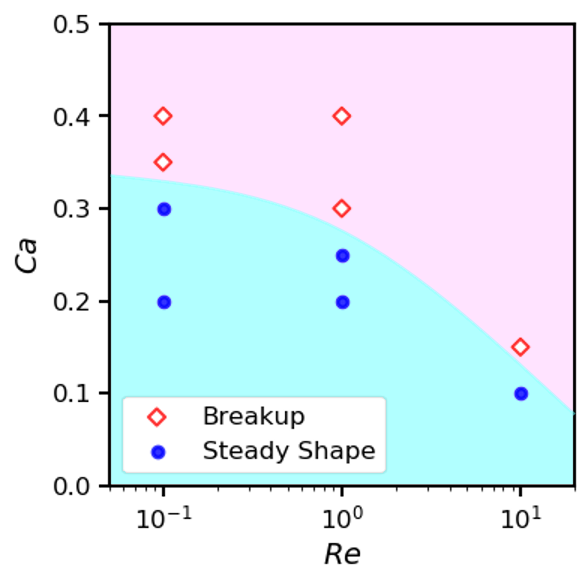} 
\caption{\label{fig:break_3d_phase} Phase diagram of 3D droplets states under different $Re$ and $Ca$ when $\alpha=\lambda=1$, $H=4R_0$, $L=32R_0$, $W=8R_0$.}
\end{figure}
Due to highly computational cost in 3D,
we only consider a moderate confinement $H/R_0=4$
and perform a group of simulations to draw
a phase diagram in the plane of $Ca$ and $Re$, as shown in Fig.~\ref{fig:break_3d_phase}. The size of the simulation box is $L=32R_0$, $W=8R_0$, $H=4R_0$. As in 2D case, the critical $Ca_c$ decreases with increasing $Re$ in 3D,
as shown in in Fig.~\ref{fig:break_re_confb}.
However, the critical capillary number $Ca_c$ in 3D case is significantly smaller than that of 2D case.

\subsection{Water droplet in air flow}
\label{sec:result_waterair}
As one specific application, we employ our method to predict the breakup of a water droplet in shear flow of air. The critical capillary number or the shear rate determined is helpful to design an effective atomization device.  Actual physical properties of water and air around 20$^{\circ}$C are adopted: $\rho_d=998.2\;kg\cdot m^{-3}$, $\mu_d=1.0087\times 10^{-3}\;Pa\cdot s$ and $\rho_c=1.205\;kg\cdot m^{-3}$, $\mu_c=1.81\times 10^{-5}\;Pa\cdot s$ are set for water (dispersed) phase and air (continuous) phase, respectively; surface tension coefficient $\sigma =72.75\times 10^{-3}\;N\cdot m^{-1}$ is set for the water-air interface.

\begin{figure} 
\centering
\includegraphics[width=0.6\textwidth]{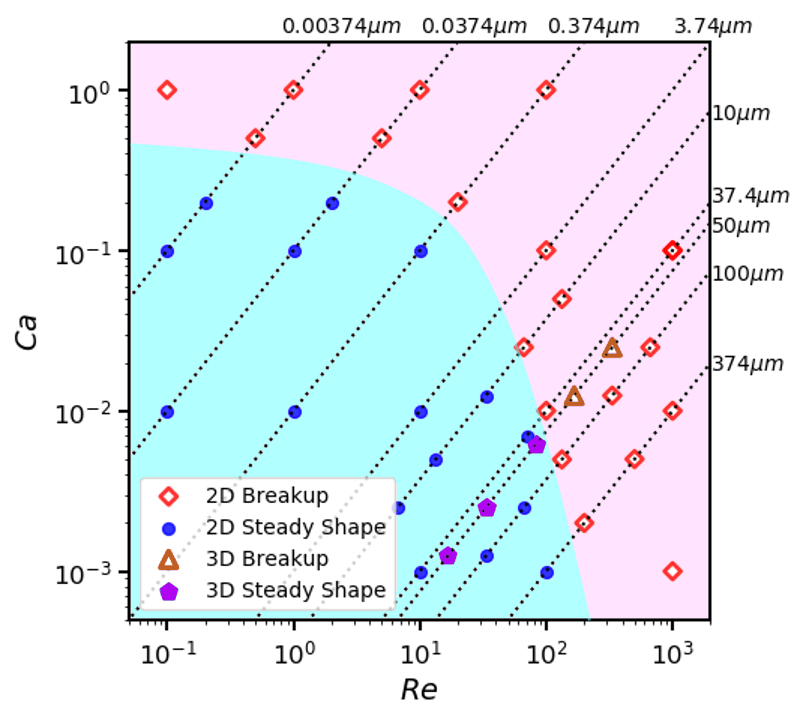} 
\caption{\label{fig:break_waterair}Diagram for states of water droplets in air shear flow under different $Re$ and $Ca$ when $H=16R_0$, $L=16R_0$.}
\end{figure}
We perform a relative large range of Reynolds numbers
and depict a phase diagram on the plane of $Re$ and $Ca$ in logarithmic-logarithmic scales on Fig.~\ref{fig:break_waterair}.
This allows us to connect the results with the same droplet size 
and observe its behavior while changing $Re$ and $Ca$.
Points on each dotted line represent the droplet of the same radius, as marked in the figure.
For example, we have a line of dynamics for the droplet with $R_0=10\mu m$ under shear rates of $1\times 10^6s^{-1}$, $2\times 10^6s^{-1}$, $5\times 10^6s^{-1}$, $1\times 10^7s^{-1}$, $2\times 10^7s^{-1}$; another line of dynamics for the droplet with $R_0=100\mu m$ under shear rates of $5\times 10^4s^{-1}$, $1\times 10^5s^{-1}$, $2\times 10^5s^{-1}$, $5\times 10^5s^{-1}$, $1\times 10^6s^{-1}$. 
Furthermore,  we observe that if the $Re$ is on the order of $100$, the critical $Ca$ for breakup is very sensitive to $Re$.
We also perform a group of 3D simulations for a droplet with $R_0=50 \mu m$ under shear rates of $1\times 10^5s^{-1}$, $2\times 10^5s^{-1}$, $5\times 10^5s^{-1}$, $1\times 10^6s^{-1}$, $2\times 10^6s^{-1}$. The 3D results for the critical point of breakup is close to that of the 2D results.


\section{Conclusions and disucssions}
\label{sec:conclusion}

In this study, we employed a multi-phase SPH method to simulate droplet deformation and breakup subjected to a simple shear flow in an extensive range of physical parameters. We performed both 2D and 3D simulations and validated them by benchmarks: transient deformations and steady shapes of droplets are compared with previous simulations, analytical derivations and experimental data. These results indicate that the method is reliable to simulate droplet dynamics in general. We wish to emphasize the convenience of SPH method in simulating multi-phase problems, as we can leverage on its Lagrangian nature and differentiate different phases by particle species. In addition, the algorithm and data structure for 2D and 3D simulations have tiny difference and therefore, it is a simple task to extend the code from 2D to 3D. 
Economical 2D simulations allow us to investigate a wide range of physical parameters in five dimensions, which serve as a guide to 3D realistic situations. 
From the results, we come to the following conclusions.

(1) A larger Reynolds number $Re$ or capillary number $Ca$ leads to a more considerable deformation of the droplet. The transient and steady-state deformations of the droplet in our study are in good agreement with the previous studies but beyond their time limits~\cite{sheth1995EffectsInertiaDeformation, zhou1993FlowSuspensionsChannels}.

(2) Under low Reynolds number ($Re=0.1$), a stronger confinement due to the walls enhances the steady-state deformation in both 2D and 3D simulations. When the walls are separated further apart, the Taylor deformation parameter is almost linear with respect to $Ca$.
The influence of confinement on the deformation of a droplet has been studied by Shapira and Haber by a first-order analytical solution based on Lorentz's reflection method. They proved that the walls do not influence the shape of deformed droplet but increases the deformation magnitude with a term of order $\left ( R_0/H \right ) ^3$~\cite{shapira1990LowReynoldsNumber}.
The experiment data of Sibillo et al. illustrate satisfactory agreement with the predictions of Shapira and Haber except for the droplet being within a small gap, where the reflection analysis is expected to fail~\cite{sibillo2006DropDeformationMicroconfined}.
Our 3D simulation results resemble the whole set of experiment data even when the droplet is within the small gap,
which suggests the method as an applicative tool for more realistic situations in microfluidics.

(3) The effects of wall confinement on the critical capillary number $Ca_c$ are not universal under different $Re$. When $Re=0.1$, a closer gap of walls reduces $Ca_c$. This is because a closer gap of walls increases the deformation as described above. But when $Re$ is larger, the relation between $Ca_c$ and the confinement ratio is unclear. From our observation, this non-monotonic relation results from an interplay of influences by the shear strength and the stability of the whole flow field. On the one hand, the shear stress transferred to the droplet from the wall is more pronounced in stronger confinement~\cite{shapira1990LowReynoldsNumber}, thus closer walls reduce the $Ca_c$. On the other hand, the narrower channel reduces the instability of the flow and restricts droplet movements, thus increases the $Ca_c$. 

(4) Under $Re=0.1$ and the range of viscosity ratio $\lambda \in \left [ 0.1,1 \right ]$, a higher $\lambda$ causes a larger deformation. The effect of $\lambda$ on $Ca_c$ is not monotonic when $\lambda > 1$ and there is a minimum value of $Ca_c$ between $\lambda=1$ and $\lambda=10$. The existence of a minimal $Ca_c$ among different $\lambda$ has also been found by previous experiment studies~\cite{karam1968deformation, grace1982dispersion}, when $\lambda$ is about $1$. 
The discrepancy between our results and the previous ones are attributed to the difference between 2D and 3D cases.
At the same $Re$, the influence of density ratio on droplet deformation is much smaller compared with that of the viscosity ratio.

(5) As an application, a phase diagram obtained by actual physical parameters of water and air is depicted to predict the magnitude of shear rate for breaking a droplet of certain size, which is helpful in the designing atomization nozzles.

\section*{Acknowledgements}
K. Wang and X. Bian acknowledge the national natural science foundation of China under grant number: 12172330.
This work is partially supported by Hangzhou Shiguangji Intelligent Electronics Technology Co., Ltd, Hangzhou, China.

\section*{Acknowledgements}
K. Wang and X. Bian acknowledge the national natural science foundation of China under grant number: 12172330.
This work is partially supported by Hangzhou Shiguangji Intelligient Electronics Technology Co., Ltd, Hangzhou, China.

 \bibliographystyle{elsarticle-num} 
 \bibliography{cas-refs}







\end{document}